\documentclass[aps,prapplied,reprint,groupedaddress,longbibliography]{revtex4-2} % You should use BibTeX and apsrev.bst for references
\usepackage{graphicx}% Include figure files
\usepackage{dcolumn}% Align table columns on decimal point

\usepackage{bm}
\usepackage{graphicx}
\usepackage[caption=false]{subfig}
\usepackage[colorlinks=true, citecolor=blue, linkcolor=blue, urlcolor=blue]{hyperref}
\usepackage{pinlabel} % pdfLatex can't load eps somehow. use pdf for figures
\usepackage{amsmath}
\usepackage{dcolumn}% Align table columns on decimal point
\usepackage{xcolor}

\usepackage{longtable}
\usepackage{amssymb}
\usepackage{natbib}
\usepackage{mathrsfs}
\usepackage[T1]{fontenc}
\definecolor{pakistangreen}{rgb}{0.0, 0.4, 0.0}
\definecolor{phthalogreen}{rgb}{0.07, 0.21, 0.14}
\newcommand{\appropto}{\mathrel{\vcenter{\offinterlineskip\halign{\hfil$##$\cr\propto\cr\noalign{\kern2pt}\sim\cr\noalign{\kern-2pt}}}}}

\newcommand{\beq}{\begin{equation}}
\newcommand{\eeq}{\end{equation}}
\newcommand{\beqr}{\begin{eqnarray}}
\newcommand{\eeqr}{\end{eqnarray}}

\newcommand{\wi}{\omega_c^{\rm I}}
\newcommand{\wia}{\omega_c^{\rm GI}}
\newcommand{\wib}{\omega_c^{\rm AI}}

\newcommand{\wE}{\omega_c^{\rm E}}

\newcommand{\RN}[1]{%
  \textup{\uppercase\expandafter{\romannumeral#1}}%
}

\usepackage{color}

\begin{document}

%\preprint{}

\title{Parasitic-free gate: A protected switch between idle and entangled states}

\author{Xuexin Xu}
%\email{x.xu@fz-juelich.de}
\affiliation{Institute for Quantum Information,
RWTH Aachen University, D-52056 Aachen, Germany}
\affiliation{Peter Gr\"unberg Institute, Forschungszentrum J\"ulich, J\"ulich 52428, Germany}
%\affiliation{J\"ulich-Aachen Research Alliance (JARA), Fundamentals of Future Information Technologies, J\"ulich 52428, Germany}

\author{M. H. Ansari}
\affiliation{Institute for Quantum Information,
RWTH Aachen University, D-52056 Aachen, Germany}
\affiliation{Peter Gr\"unberg Institute, Forschungszentrum J\"ulich, J\"ulich 52428, Germany}

\begin{abstract}
We propose a  gate to switch superconducting qubit pairs in and out of a two-body interaction. This gate uses  cross resonance driving on a tunable circuit with adjusted parameters and without accumulating residual $ZZ$ interaction for idle and interacting qubits. It is imperative that this gate does not spread errors through the quantum register. Our detailed theoretical results show that these error-free modes do not necessarily require largely tunable circuits, such as magnetic modulation of qubits or couplers. We obtain the operational gate on weakly tuneable circuits as well and show that switching between them is remarkably fast.

%We propose a protected 2-qubit gate composed of two sub-gates that switches between idling and entangling modes. These sub-gates are separated by a large energy barrier.  In superconducting circuits, this gate is derived by harnessing microwave driving and coupling modulation.  The idle mode decouples qubits and leaves them with highly localized wave functions. The entangled mode, however, makes them interact in absence of the adversarial $ZZ$ interactions.  Our theoretical results show that this gate is a promising way to switch on and off a fast and high fidelity 2-qubit interaction.
\end{abstract}


\maketitle


\section{Introduction} 
Over a few decades quantum computing has evolved from a concept~\cite{feynman1982simulating} to experiments on noisy intermediate-scale quantum processors ~\cite{preskill2018quantum,bharti2021noisy,linghu2022quantum}. In the processors, entangled qubits together search for the answer state to a computational problem and this outperforms classical computational time ~\cite{arute2019quantum,foxen2020demonstrating,wu2021strong}. In gate-based quantum processors, quantum maps are decomposed into a sequence of one and two-qubit gates. Engineering of these gates has so far targeted fast and less erroneous quantum state change. However, further developments are needed to improve gate fidelity and, as soon as gates are deactivated, idle qubits bring them within the threshold of fault-tolerant computation~\cite{harper2019fault-tolerant,chen2021exponential,noiri2022fast}.

In superconducting qubits, one of the main sources of two-qubit gate phase errors is the parasitic $ZZ$ interaction, which enforces repulsion between computational energy levels and non-computational ones.   In state-of-the-art circuits the qubit-qubit residual $ZZ$ repulsion is typically 50-100 times weaker than their coupling strength, however, this is sufficient for 1$\%$ of gate fidelity reduction over the typical gate length of $\sim 0.1~\mu s$ \cite{krinner2020demonstration,kandala2021demonstration,zhao2022quantum}.  

\begin{figure}[t]
	\centering			
\hspace{0.35in}\includegraphics[width=0.5\textwidth]{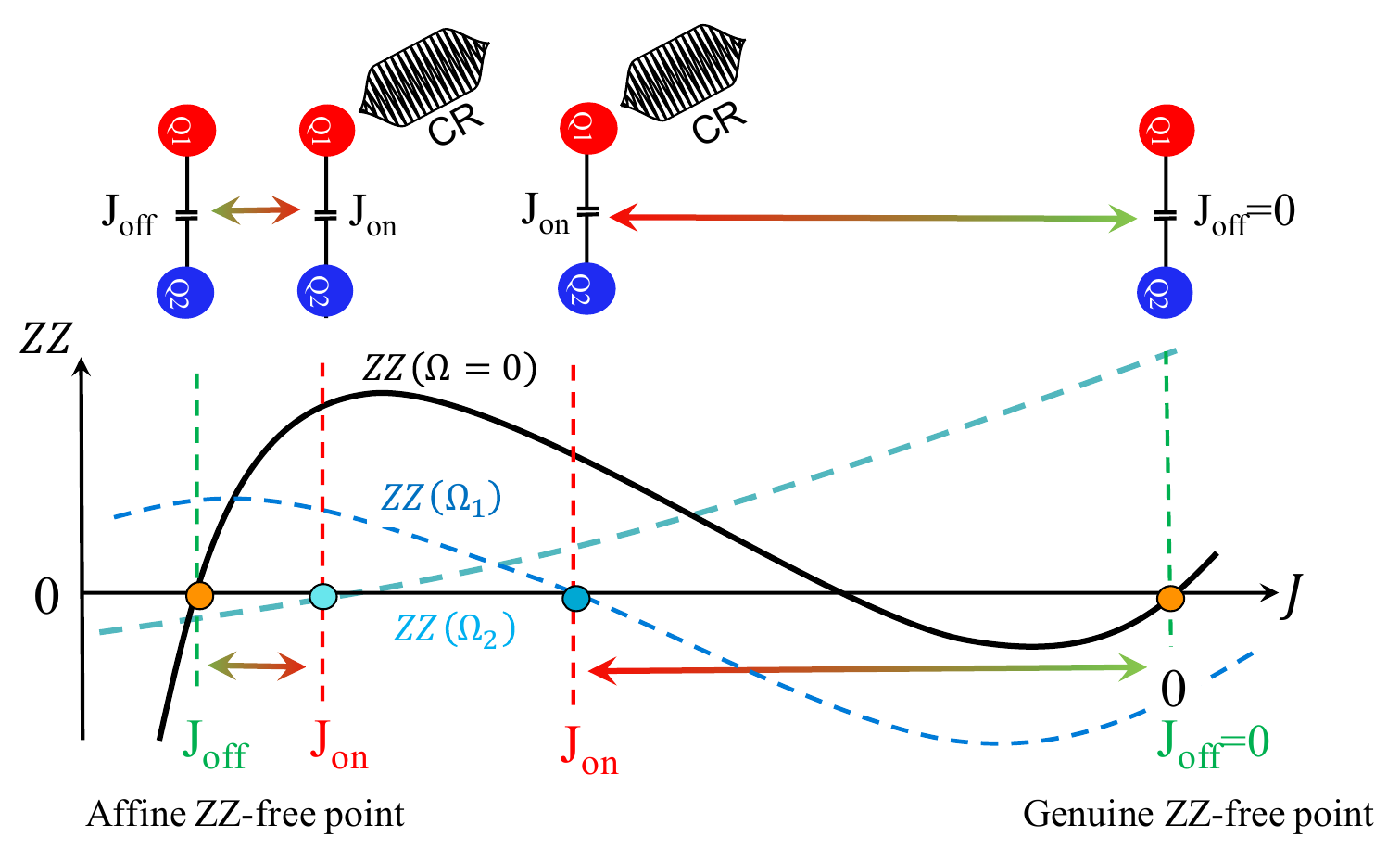}\\
		\vspace{-0.15in}
	\caption{Schematic switching PF gate between idle and entangled modes.  Net parasitic $ZZ(\Omega)$ interaction after activating CR($\Omega$) gate may vanish at certain  coupling strengths, i.e. $J_{\textup{on}}$ points. The static $ZZ$ interaction, shown as $ZZ(\Omega=0)$, has zeros at genuine and affine points. }\label{fig:diagram}
	\vspace{-0.1in}
\end{figure}

Remarkable advances for perfecting gates have been made by exactly zeroing the static $ZZ$ interaction between flux qubit and transmon~\cite{zhao2020high-contrast,ku2020suppression,jin2021implementing}, and a pair of transmons~\cite{sung2021realization,ni2021scalable}. Ungated qubits built on chip repel one another and then lead to the so-called static $ZZ$ interaction. Zeroing this interaction can take place at either a \emph{genuine} point, where qubits are decoupled, or an \emph{affine} point, where repulsions from both sides of computational levels cancel each other, see Fig.~\ref{fig:diagram}.  The distinction between genuine or affine labels, however, is basis dependent; an affine pair of qubits with frequency detuning stronger than coupling strength are effectively decoupled (with  slightly shifted frequencies) in the eigenmode basis~\cite{richermaster}.  External driving adds a new component to the total $ZZ$ strength, namely the dynamic part,  which depends on circuit parameters and driving amplitudes. Further progresses have been made recently on  perfecting the Cross Resonance (CR) gate by eliminating all residual $ZZ$ interactions \cite{xu2021zz-freedom,wei2021quantum}.  A large class of circuits under CR gates can cancel out their static $ZZ$ strength, making a net-zero $ZZ$ gate whose fidelity is only limited by qubit coherence times. 

Perfecting qubits and gates seems to be necessary steps toward error-free quantum computation, however, they are insufficient as such a computation also needs perfect idle qubits.  Deactivating a $ZZ$-free gate without readjusting circuit parameters can harm the quantum register because this eliminates the dynamic part of $ZZ$ interaction and leaves qubits with the finite static part. During the entire time that gates are not active, the seemingly idle qubits collect phase errors that over time not only grow larger but also can easily spread throughout the entire multiqubit states.

Here we introduce a new gate called the parasitic-free (PF) gate by combining a tunable circuit and CR driving. Compared to the recently implemented CR gate in a tunable coupling superconducting circuit~\cite{cai2021impact}, this gate can switch between idle (I) and entangled (E) modes without accommodating any residual $ZZ$ interaction in either mode. Figure~\ref{fig:diagram} schematically shows residual $ZZ$ interaction in the presence or the absence of CR amplitude $\Omega$  over a large domain of coupling strength. Switching mainly occurs by enabling tunability in coupling strength $J$ between qubits. In E mode, qubits are parked at a non-zero static $ZZ$ point, say at $J_{\textup{on}}$, where activating CR pulse sets its total parasitic $ZZ$ to zero, so that the  qubits purely $ZX$-interact. Deactivating external driving returns the total parasitic interaction to a non-zero static $ZZ$ strength, therefore changing coupling to  $J_\textup{off}$ helps send them to a static $ZZ$-free point, either the genuine or an affine point. 
Our theoretical results show that the two modes can be nearby so that tuning does not necessarily require supplying large difference between $J_{\textup{on}}$ and $J_\textup{off}$,  see the tuning in the vicinity of the affine point in Fig.~\ref{fig:diagram}.  We find a large class of circuits in which weakly tunable circuits can accommodate both modes. An interesting realization of such circuits can be the recent proposal of magnetic-free couplers between two qubits that are weakly tuned by mutual inductive coupling to an external inductance~\cite{chavez-garcia2022weakly}.

\vspace{-0.2in}
\section{Principles} \vspace{-0.1in}\label{sec:principle}
The $ZZ$ parasitic-free (PF) gate is a reversible operation that requires modulating a circuit parameter to switch it between two modes: undriven idle (I) mode, and driven entangled (E) mode. Qubits at I mode are free from residual $ZZ$ interaction.  To bring the qubits to desirably interact at E mode, they are first brought to couple with finite static $ZZ$, then a microwave pulse drives them so that they start to interact with a desired $ZX$-type coupling at the same time they are liberated from net parasitic $ZZ$ interaction.  This gate combines two important features: 1) by filtering out parasitic interactions in presence or absence of external driving, high state fidelity can be achieved on both modes, and 2) it safely provides faster as well as higher fidelity 2-qubit gate by combining circuit tunability with external driving. 

\begin{figure}[t]
	\centering			
\hspace{0.25in}\includegraphics[width=0.4\textwidth]{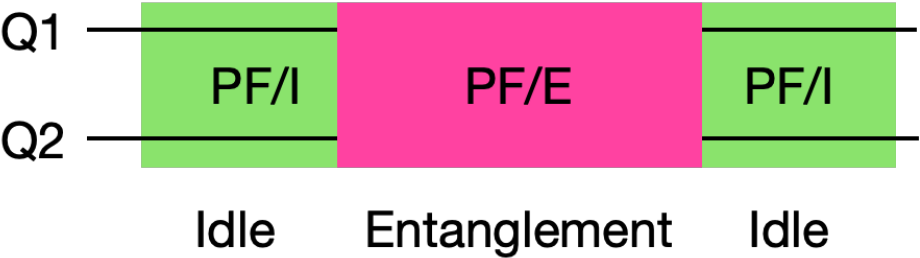}\put(-237.5,48){(a)}\\
	\includegraphics[width=0.48\textwidth]{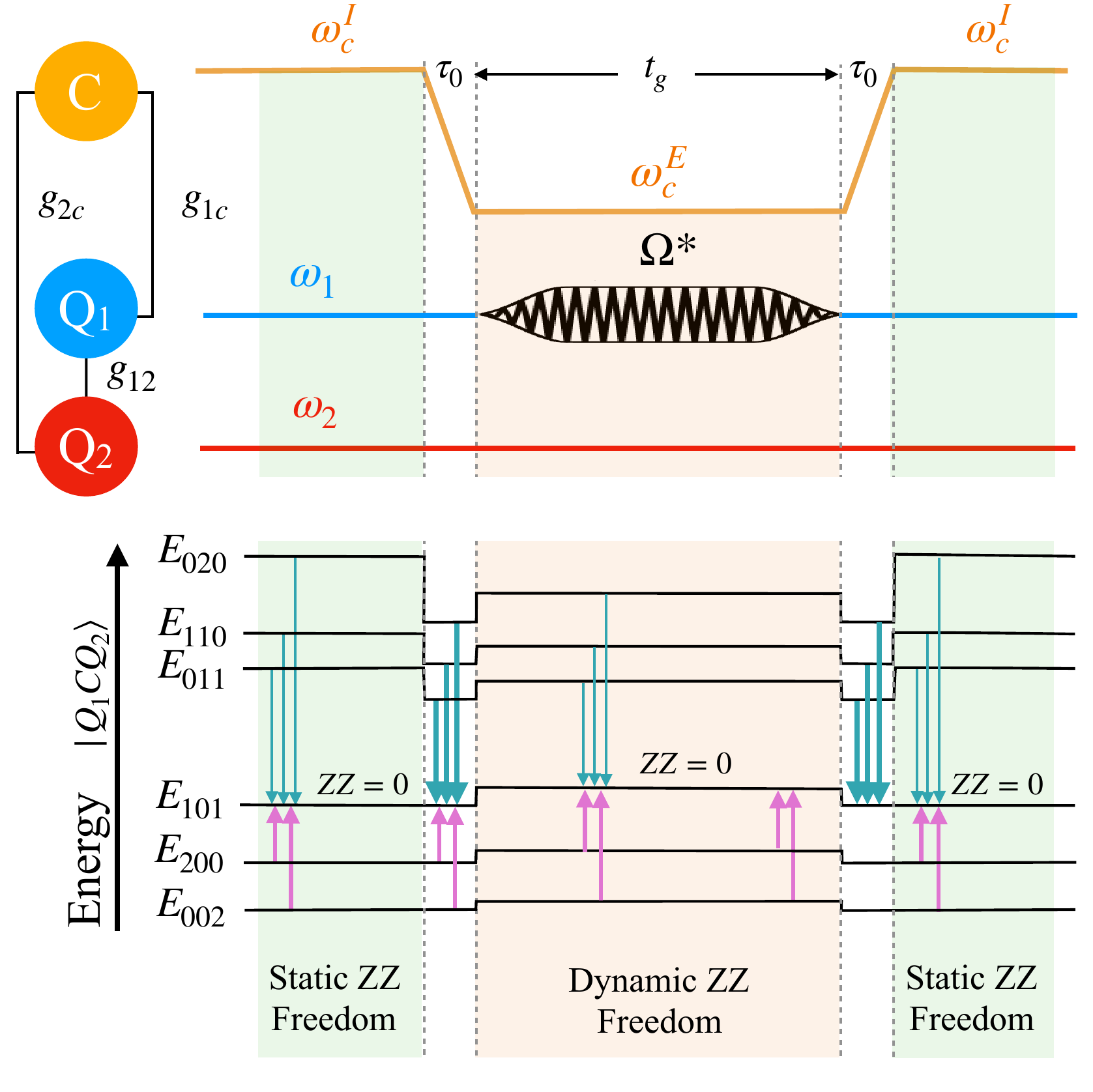}\put(-250,235){(b)}\put(-250,125){(c)}\\
	\vspace{-0.15in}
	\caption{(a) The PF gate at idle mode is active during PF/I operation box (green) and at  entangled mode during PF/E operation box (pink).  (b) Left: PF gate circuit; Right: PF gate timing and components. $t_g$ is the duration of the entangled mode with $\tau_0$ being the rise/fall time from idle to entangled modes}; (c) The energy diagrams at idle and entangle modes -- vertical arrows shows level repulsions. \label{fig:cird}
\end{figure}

Figure~\ref{fig:cird}(a) shows the schematic of the PF gate that switches qubits Q1 and Q2  into either one of the following two modes:
\begin{description}
	\item[PF/I mode] represented by a green box--- where qubits are $ZZ$-free by tuning a circuit parameter, such as mutual coupling strength, and this hibernates their initial state $|Q_1 Q_2\rangle$ as long as they are at the idle (I) mode, 
	
	\item[PF/E mode] represented by a pink box--- where qubits are coupled after changing circuit parameter and then applying microwave driving on qubits. These make them interact only by $ZX$ in absence of unwanted $ZZ$ interactions as long as the microwave is on.    
\end{description}

The switching operation can be accomplished in either frequency-tunable qubits, tunable coupler between qubits, or combining both. The gate principles are universal for all types of qubits and both harmonic and anharmonic couplers. However, there is a practical preference to use a tunable coupler rather than tunable qubits, since the latter is proved to slightly suffer from rather lower coherence times that may degrade gate performance~\cite{hutchings2017tunable}.  

Let us consider a circuit with qubits Q1 and Q2 coupled via the coupler C and denote their quantum states as $|Q_1, C, Q_2\rangle $.  Schematic circuits can be seen in Fig.~\ref{fig:cird}(b), where Q1 and Q2 interact directly by $g_{12}$ and indirectly by individual couplings to C with coupling strengths $g_{1c}$ and $g_{2c}$.  In principle, qubit and coupler Hamiltonians are similar since  coupler can be considered as a third qubit, i.e. $H_i=\omega_i (n_i) \hat{a}_i^\dagger \hat{a}_i+\delta_i  \hat{a}_i^\dagger \hat{a}_i^\dagger \hat{a}_i \hat{a}_i /2$, with $\hat{a}_i$ ($\hat{a}^\dagger_i$) being annihilation (creation) operator, $\omega_i$  frequency, $\delta_i$ anharmonicity, and  $i=1,2,c$. We can write circuit Hamiltonian as  $H=\sum_{i=1,2,c} H_i + \sum_{i\neq j} g_{ij} (\hat{a}^\dagger_i+\hat{a}_i)(\hat{a}_j^\dagger+\hat{a}_j)$ with $g_{ij}$ being coupling strengths. In the situation where qubits are far detuned from coupler, $|\omega_{1/2}-\omega_c|\gg |g|$, namely the dispersive regime,  the total Hamiltonian can be perturbatively diagonalized in higher order of $g/|\omega_{1/2}-\omega_c|$. However, it is important to emphasize that quantum processors can operate beyond the dispersive regime, see Ref.~\cite{ansari2019superconducting}. 

By summing over coupler states and transforming the Hamiltonian into a block diagonal frame \cite{bravyi2011schrieffer--wolff}, one can simplify it as an effective Hamiltonian in the computational Hilbert space of two qubits \cite{magesan2020effective}. This simplification reveals that the two qubits interact only by a $ZZ$ interaction, which is usually considered unwanted and always-on as long as energy levels are not shifted. The computational part of effective Hamiltonian in its eigenbasis, namely `dressed basis', is 
\beq
H_{\rm eff}=-\tilde{\omega}_1 \hat{Z}\hat{I}/2- \tilde{\omega}_2 \hat{I}\hat{Z}/2 + \zeta_s \hat{Z}\hat{Z}/4
\label{eq. Heff} 
\eeq 
with $\tilde{\omega}_i$ being qubit frequency in tilde dressed basis and  $\zeta_s =\tilde{E}_{11}-\tilde{E}_{01}-\tilde{E}_{10}+\tilde{E}_{00}$ being the static level repulsion coefficient in absence of external driving.  As long as qubits are not externally driven Eq.~(\ref{eq. Heff}) describes the circuit quantum electronics to acceptable accuracy.

Externally driving Q1, namely \emph{control} qubit, at the frequency of Q2, namely \emph{target} qubit on superconducting circuits, introduces the operator driving Hamiltonian:  $H_{dr} = \Omega \cos (\tilde{\omega}_2 t) (\hat{a}_1^\dagger + \hat{a}_1)$.  This operator in states representation can be written as $\sum_{n_c,n_2} ( |0, n_c, n_2\rangle \langle 1, n_c, n_2| + |1, n_c, n_2\rangle \langle 2, n_c, n_2| +\cdots + H.c.)$ and induces that external driving triggers some transitions on target qubit. In the frame co-rotating with driving pulse, ignoring highly excited levels simplifies the Hamiltonian in the leading order of $\Omega $: 
 \beqr
 \label{eq. Horig}
  H_d  &&= \lambda_1  \Omega \left(  |000\rangle \langle 001| -  |100\rangle \langle 101|\right)   \\ &&
+ \lambda_2    \Omega  \left(  |000\rangle \langle 010| -   |100\rangle \langle 110| \right)  \nonumber \\ &&
 + \lambda_3    \Omega \left(  |001\rangle \langle 002| -   |101\rangle \langle 102| \right)     \nonumber \\ &&  
 +  |001\rangle \left(  \lambda_4  \Omega \langle 011| +\lambda_5  \Omega \langle 200| \right)   \nonumber \\ &&
 +  |010\rangle \left( \lambda_6  \Omega \langle 200|+ \lambda_7  \Omega \langle 011| +\lambda_8  \Omega \langle 020| \right)      \nonumber \\ &&
 +  \left[  \lambda_{9}   \Omega | 011\rangle + \lambda_{10}  \Omega |200\rangle + \lambda_{11}  \Omega | 002\rangle \right]  \langle201|  + H.c.   \nonumber
 \eeqr
In Appendix \ref{app.d26} we briefly explain how one can derive this Hamiltonian~(\ref{eq. Horig}) and evaluate all $\lambda$'s in the leading $g^2$ order.

\begin{figure}[h!]
	\centering
	\includegraphics[width=0.4\textwidth]{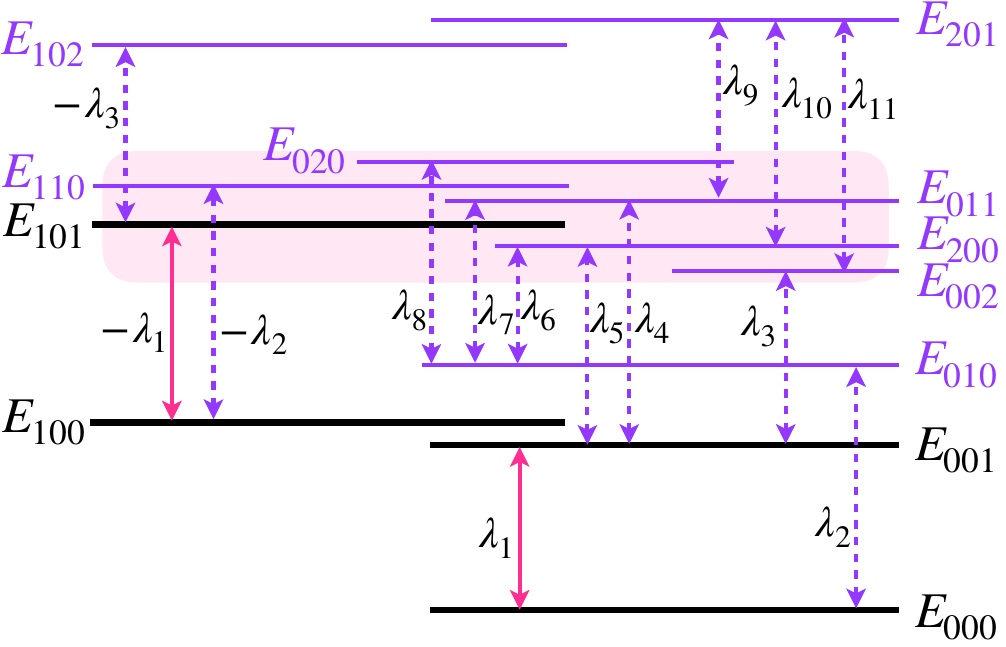}
	\vspace{-0.1in}
	\caption{Energy levels $E_{q_1,c,q_2}$ and microwave-driven transitions in a frame co-rotating with the microwave pulse.  Double arrowed solid (dashed) lines show computational (noncomputational) transitions. Shaded area shows near $E_{101}$ zone.}\label{fig:diag}
\end{figure}

In the Hamiltonian~(\ref{eq. Horig}) we dropped driven unwanted interactions that are experimentally removable, such as $ |000\rangle \langle 001| + |100\rangle \langle 101|$ and $  |001\rangle \langle 002| +   |101\rangle \langle 102| $, which acts as a single-qubit gate on Q2. In practice, a secondary simultaneous extrenal pulse should be applied on the target qubit with certain characteristics to eliminate these unwanted interactions, such as $IX$, $ZY$, and $IY$. Moreover, by either echoing the pulses or by applying virtual $Z$ gate one can eliminate $ZI$ component. The transitions listed in Eq. (\ref{eq. Horig}) are schematically shown in Fig.~\ref{fig:diag} for typical circuit frequencies used in Fig.~\ref{fig:cird}(b).

Evidently in the dispersive regime  by mapping the Hamiltonian (\ref{eq. Horig}) on computational subspace one can obtain a microwave assisted  part for $ZZ$ interaction between qubits~\cite{magesan2020effective,malekakhlagh2020first-principles}, denoted here by $\zeta_{d}$.  This indicates that total $ZZ$ interaction in presence of driving pulse is $\zeta=\zeta_s + \zeta_d$. 

Let us now supply further details about the static part. The coupler between two qubits can be a harmonic oscillator, such as a resonator, or another qubit with finite anharmonicity. The perturbative analysis of a  harmonic coupler shows that it supplies the effective $ZZ$ coupling $\zeta_s^{(1)}$ between two qubits, which depends on circuit parameters as shown in Eq. (\ref{eq.zeta}). A finite anharmonicity $\delta_c$ for the coupler will add the correction $\zeta_s^{(2)}$.  In Eq.~(\ref{eq.zeta}) we shows both parts in $O(g^4)$:
\begin{eqnarray}
 \zeta&=&\zeta_{\rm s}^{\rm (1)}+\zeta_{\rm s}^{\rm (2)}+\zeta_{d}, \nonumber \\
 \zeta_{\rm s}^{\rm (1)}& =& \frac{ 2g_{\rm eff}^2\left( \delta_1+\delta_2\right) }{(\Delta_{12}-\delta_2)(\Delta_{12}+\delta_1)}, \nonumber \\
 \zeta_{\rm s}^{\rm (2)}&=&\frac{8(g_{\rm eff}-\chi g_{12})(g_{\rm eff}- g_{12})}{\Delta_1+\Delta_2-\delta_c}\label{eq.zeta},
\end{eqnarray} 
with $\Delta_{12}=\omega_1-\omega_2$, $\Delta_q=\omega_q-\omega_c$, $\chi=\delta_c/(\Delta_1+\Delta_2)$, and the effective coupling between two qubits is 
\begin{equation}\label{eq.geff}
g_{\rm eff}=g_{12}+\frac{g_{1c}g_{2c}}{2}\sum_{q=1,2}\left(\frac{1}{\Delta_q}-\frac{1}{\Sigma_q}\right),
\end{equation}
with $\Sigma_q=\omega_q+\omega_c$.

At the entangled mode of the PF gate, firstly qubits are coupled by changing circuit parameters and therefore a non-zero static $ZZ$ interaction is expected to show up between qubits. A cross resonance pulse is then assisted so that $ZX$ interaction is supplied between qubits. Let us denote the strength of this coupling with $\alpha_{ZX}$.  From Eq.~(\ref{eq. Horig}) one can determine it in the leading perturbative order $\alpha_{ZX} \sim \lambda_1\Omega$ which is in agreement with experiment in weak $\Omega$ regime \cite{sheldon2016procedure}. Any further nonlinearity can be studied in higher orders.  The $ZX$ interaction transforms quantum states by the operator $\hat{U}=\exp (  2\pi i  \alpha_{ZX} \tau  \hat{Z}\hat{X}/2)$ during the time $\tau$ that external driving is active.  In order to perform a typical $\pi/2$ conditional-rotation on the second qubit, i.e. $ZX_{90}$, the two-qubit state must transform by $\exp (i (\pi/2) \hat{Z}\hat{X}/2) $. This indicates that external driving should be switched on for $\tau=1/4\alpha_{ZX}$.  Therefore the stronger $\alpha_{ZX}$ is the shorter time performing the gate takes.  Needless to say that during the whole time the driving is active, this interaction is accompanied by the driving-assisted parasitic interaction $\zeta_{d}$. However, there is a good chance that one can find a large class of parameters at which the total $ZZ$ interaction vanishes. Interestingly we show in the next section that strengthening $\alpha_{ZX}$ can be found by modulating both coupling strength between qubits and the external driving amplitude. This improves the gate performance not only by zeroing parasitic interactions but also by making the gate much faster. 

In the following sections, we will discuss a detailed analysis of several circuit examples and show the performance of the PF gate on them.

\vspace{-0.2in}
\section{Examples of the PF gate} \vspace{-0.1in}\label{sec:realization}

\begin{figure}[h]
	\centering
	\includegraphics[width=0.49\textwidth]{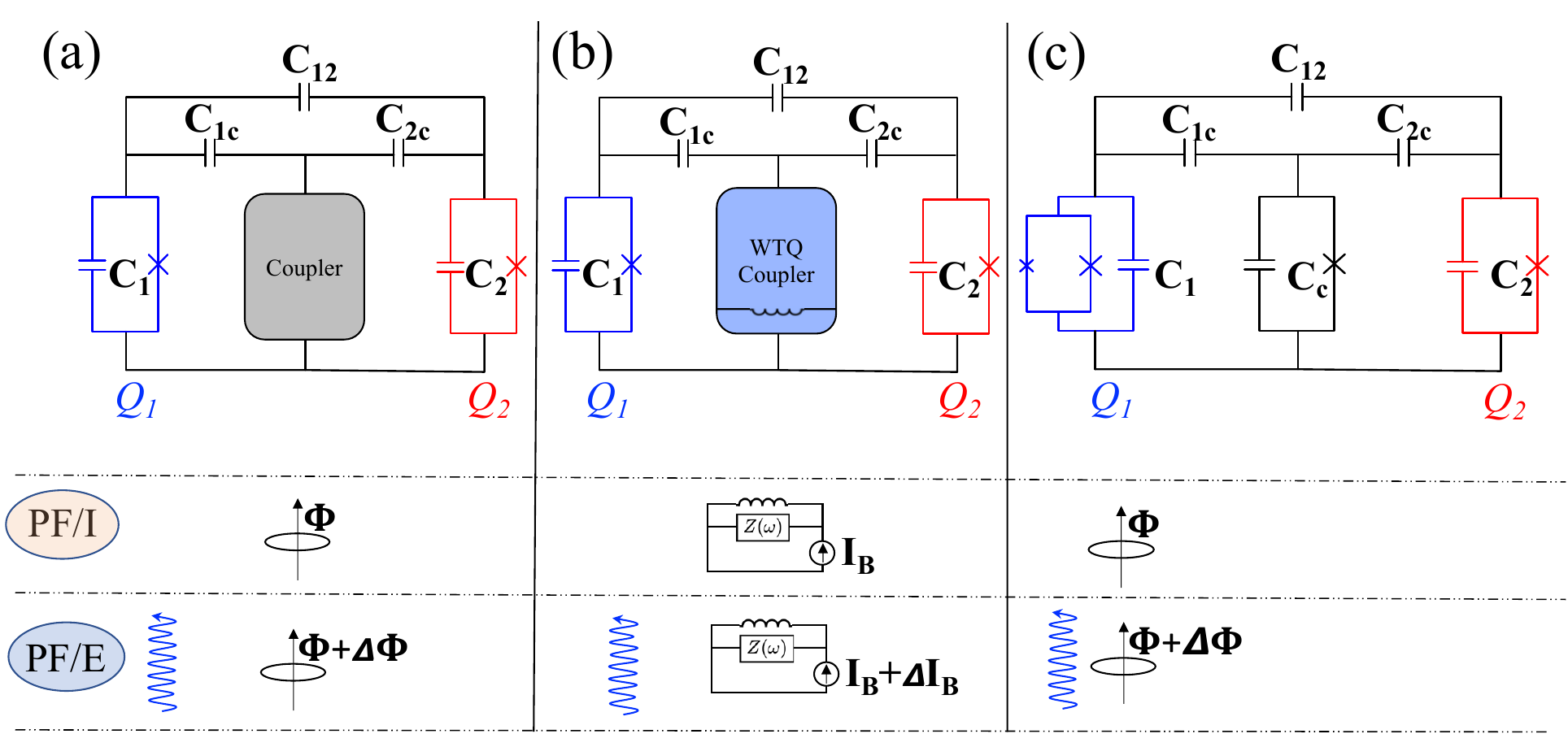}
	\vspace{-0.2in}
	\caption{Circuits with two qubits interacting via a coupler for implementing PF gate, contains (a) wide frequency-tunable coupler (b) weakly tunable coupler, and (c) a tunable frequency qubit. Lower panels show how PF gate on each circuit switches between I and E modes.}
	\label{fig:circuit}
\end{figure}

Switching between I and E modes requires a change in circuit parameter before external driving is activated.  In a circuit with two qubits and a coupler there are different possibilities for selecting which one is tuned by circuit parameter modulation and which one is driven externally. Figure~\ref{fig:circuit} shows three possible examples based on superconducting qubits. In circuit (a) two fixed frequency qubits Q1 and Q2 are coupled by a tunable coupler, which can be another qubit with flux-tunable frequency, so that as one can see in Eq.~(\ref{eq.geff}) changing the flux modulates effective coupling strength between qubits. In circuit (b) two qubits are coupled to a weakly tunable qubit (WTQ), which enables its frequency to be tuned in a small range by manipulating inductive coupling~\cite{chavez-garcia2022weakly}. Circuit (c) is different as it consists of a flux-tunable  Q1 coupled to fixed-frequency Q2 via a fixed-frequency coupler, but it suffers from a rather limited qubit coherence time. For all the three circuits, the E mode can be assisted by externally driving Q1. It is worth mentioning that there are other microwave-activated approaches to implement quantum gates, for instance, imposing additional pulses in circuit (a) has recently proved useful to execute multiqubit gate experiment~\cite{kim2021highfidelity,baker2021single}.

Here we study circuit (a) with tunability in the coupler, but it does not mean the coupler must be an asymmetric transmon. In this circuit PF/I mode is obtained by tuning the coupler to the frequency $\wi$, and in PF/E mode it is tuned at $\wE$ accompanied by a CR drive, as shown in Fig.~\ref{fig:cird}(b).

To test the performance of the PF gate we numerically study seven sample devices all parametrized on the circuit of Fig.~\ref{fig:circuit}(a).  These devices are listed in Table \ref{tab:device}. The capacitive direct coupling $g_{12}$  are grouped in 3 values, the weakest for device 1, intermediate for devices 2, 3, 4, 7, and the strongest in devices 5 and 6. Among devices in the intermediate $g_{12}$ group device 3 has stronger coupler anharmonicity, while devices 2 and 4 have similar coupler anharmonicity, yet in device 4 qubit anharmonicity is stronger. Specifically, device 7 stays out of straddling regime where $|\Delta|>|\delta|$ and the tunable coupler has positive anharmonicity.
In the group of devices 5 and 6 with the strongest $g_{12}$, the qubit-qubit detuning frequency is stronger compared to all other devices with the difference that device 5 is on a hybrid circuit by combining a transmon and a Capacitively Shunted Flux Qubit (CSFQ) while device 6 is on a transmon-transmon circuit. We consider universal qubit-coupler coupling strength $g_{1c}/2\pi=g_{2c}/2\pi=95$~MHz parked at $\omega_c=4.8$ GHz for all devices.

\begin{longtable}[t]{@{\extracolsep{\fill}}ccccccc@{}}
	\caption{Device parameters.}
	\label{tab:device}
	\endfirsthead
	\endhead
	\hline\hline
	\centering
	& $\omega_1/2\pi$ &$\omega_2/2\pi$ &$g_{12}/2\pi$ & $\delta_c/2\pi$ & $\delta_1/2\pi$ &$\delta_2/2\pi$ \\ 
	& (GHz) &(GHz) & (MHz) & (MHz) & (MHz) & (MHz)\\
	\hline
%	1 & 4.3 & 4.20 & 3.76 &$-100$ &$-250$ &$-250$\\
	1& 4.25 & 4.20 & 3.76 &$-100$ &$-250$ &$-250$\\
	2 & 4.25& 4.20 & 6.48 &$-100$ &$-250$ &$-250$\\
	3 & 4.25 & 4.20 & 6.48 &$-200$ &$-250$ &$-250$\\
	4 & 4.25 & 4.20 & 6.48 &$-100$ &$-320$ &$-320$\\
	5& 4.00 &4.20&9.48&$-100$&$~~500$&$-250$\\
	6& 4.40 &4.20&9.48&$-100$&$-320$&$-320$\\
	7& 4.50 &4.20&6.48&$+200$&$-250$&$-250$\\
	\hline\hline
\end{longtable}

We evaluate the static $ZZ$ interaction using the parameters listed in Table~\ref{tab:device}. We take three different approaches  for our evaluations. In one approach we numerically diagonalize the Hamiltonian~(\ref{eq.hcoupler}) in a large Hilbert space. We tested these results with yet another numerical formalism proposed recently in Ref.~\cite{li2021non-perturbative}, namely the Non-Perturbative Analytical Diagonalization (NPAD) method. These two methods give rise to the same result as plotted in Fig. (\ref{fig:staticzz}) in Appendix \ref{app compare}.  In the same plot we also present the second order Schrieffer-Wolff perturbative results as SWT, which turns out to be consistent with the numerical results only when the coupler frequency is tuned far away from qubits.

\begin{figure}[tp]
	\centering
	\includegraphics[width=0.48\textwidth]{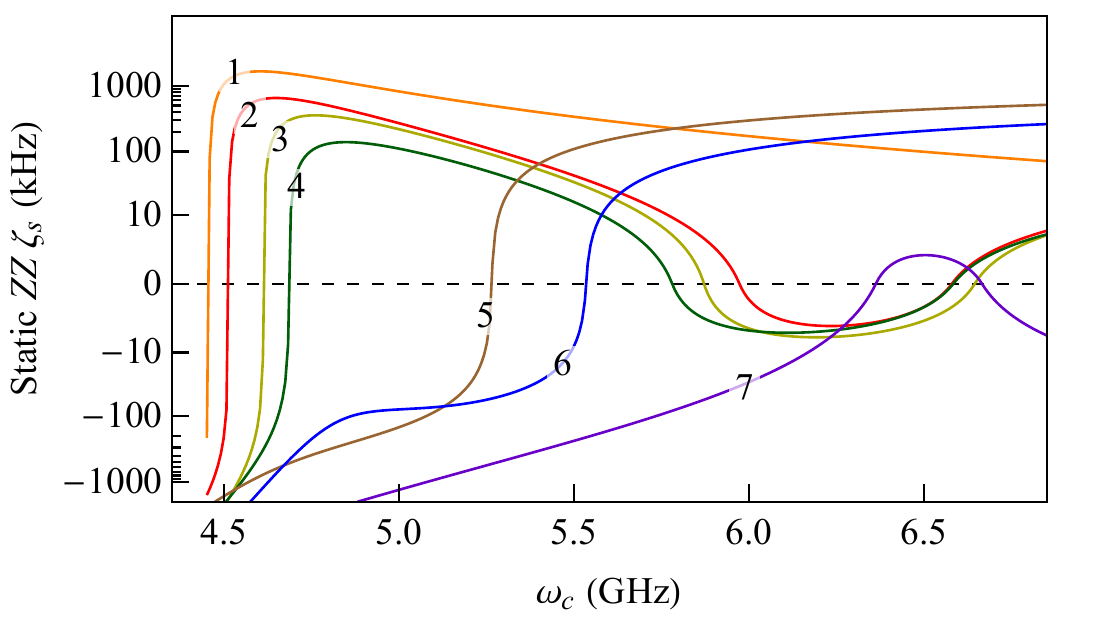}
	\vspace{-0.3in}
	\caption{Numerical static $ZZ$ strength in the seven devices listed in Table~\ref{tab:device} versus coupler frequencies. }
	\label{fig:coupler2zz}
\end{figure}

Figure~\ref{fig:coupler2zz} shows numerical values for the static $ZZ$ interaction at different coupler frequency $\omega_c$ for the seven devices listed in Table \ref{tab:device}.  One can see that all devices possess at least one zero-$ZZ$ point. This will be further discussed below.

\vspace{-0.2in}
\subsection{The idle mode}

In the circuit of Fig.~\ref{fig:circuit}(a), if the coupler frequency is far detuned from qubits, there may exist a certain coupler frequency at which the effective interaction between qubits vanishes, i.e. $g_{\rm eff}=0$. Using Eq.~(\ref{eq.geff}) one can easily find the answer. However, if the coupler frequency is closer to qubits, $\zeta_2^{(2)}$ induced by the anharmonic coupler becomes comparable with $\zeta_2^{(1)}$, making it possible to achieve static $ZZ$ freedom as shown on the far left of Fig.~\ref{fig:coupler2zz}. These two static $ZZ$ freedoms correspond to two types of idle modes, here we call the far-right point with $g_{\rm eff}=0$ in Fig.~\ref{fig:coupler2zz} as a genuine $ZZ$-free point, and the far left point in Fig.~\ref{fig:coupler2zz} as an affine $ZZ$-free point. There is also a third type of $ZZ$ zeroness which stays between genuine and affine points, however, it always shows up accompanied with at least one of the genuine/affine $ZZ$-free points, we treat it as a trivial solution and will not further discuss it. 

\paragraph*{Genuine idle (GI) mode:} Let us first study the genuine $ZZ$-free point with effective coupling $g_{\rm eff}=0$. As discussed in Ref.~\cite{yan2018tunable} in a circuit couplings between interacting elements are frequency dependent. The qubit-coupler interaction strengths $g_{1c}$ and $g_{2c}$, denoted in Fig.~\ref{fig:cird}(b), can be rewritten in terms of capacitances shown in the analogue circuit of Fig.~\ref{fig:circuit}(a). The  relation between two sets of parameters can be approximated as follows:   $g_{ic}\approx \alpha_{i}  \sqrt{\omega_i \omega_c}$ and $g_{12}\approx \alpha_{12} \sqrt{\omega_1 \omega_2}$ for the qubit label $i=1,2$, with $\alpha_{i}=C_{ic}/2\sqrt{C_i C_c}$ and $\alpha_{12}=(C_{12}+C_{1c} C_{2c} / C_{c}) /2\sqrt{C_1 C_2}$, more accurate derivation can be found in Refs.~\cite{didier2018analytical,ku2020suppression}. By substituting these relations into Eq.~(\ref{eq.geff})  one can find the so-called genuine idle coupler frequency $\wia$ at which qubits are effectively decoupled:
\begin{equation}\label{eq.offwa}
\wia= \frac{\omega_1+\omega_2}{2\sqrt{1-2\alpha_1\alpha_2/\alpha_{12}}}.
\end{equation}

By substituting $\wia$ in Eq.~(\ref{eq.zeta}) the static $ZZ$ interaction turns out to have a small offset $8g_{12}^2\delta_c/(\omega_1+\omega_2-2\wia)^2$. Usually, this offset in the dispersive regime is only a few kilohertz due to the inaccuracy of second-order perturbation theory used to derive Eq.~(\ref{eq.zeta}). For example, in the circuit used in Ref.~\cite{sung2021realization} $\alpha_{1/2}\sim10\alpha_{12}\sim0.02$ and $\omega_{1/2}\sim4$ GHz, the offset is found approximately $-2$ kHz. This is the main reason we do not limit our analysis in the rest of the paper into perturbation theory and instead, we take a more accurate approach of numerical Hamiltonian diagonalization. Further comparison can be found in Appendix~\ref{app compare}. The numerical result shows that $\wia$ is slightly shifted from Eq.~(\ref{eq.offwa}) by a few MHz. This difference for the seven circuits is given in Table~\ref{tab:offa}.

\vspace{0.2in}
\begin{longtable}{@{\extracolsep{\fill}}cccccccc@{}}
	\caption{Numerical and perturbative coupler frequency $\wia$.}
	\label{tab:offa}
	\endfirsthead
	\endhead
	\hline\hline\\[-2ex]
	\centering
	
	$\wia/2\pi$ (GHz)&1 & 2& 3&4 &5 &6 & 7\\
		\hline\\[-1.5ex]
	Numeric & NA&6.577 &6.643 &6.577&5.261 & 5.532& 6.674 \\[0.5ex]
	Eq.~(\ref{eq.offwa}) & NA &6.522 &6.522 &6.522 &5.278&5.536&6.715\\[0.5ex]
	\hline\hline
\end{longtable}

\paragraph*{Affine idle (AI) mode:} When the coupler frequency is closer to qubits, effective coupling $g_{\rm eff}$ is strengthened such that $g_{\rm eff}\gg g_{12}$. In this case by solving $\zeta^s=0$ we have the following perturbative idle coupler frequency:
\begin{equation}\label{eq.offwb}
\wib \approx \frac{\omega_1+\omega_2-\delta_c}{2}-\frac{2(\Delta_{12}-\delta_2)(\Delta_{12}+\delta_1)}{\delta_1+\delta_2}
\end{equation}

Table~\ref{tab:offb} compares the numerical simulation of $\wib$ with the perturbative results, note that the perturbative solution beyond the dispersive regime has been ignored.  
\begin{longtable}{@{\extracolsep{\fill}}cccccccc@{}}
	\caption{Numerical and perturbative coupler frequency $\wib$.}
	\label{tab:offb}
	\endfirsthead
	\endhead
	\hline\hline\\[-2ex]
	\centering
	
	$\wib/2\pi$ (GHz)&1 & 2& 3&4 &5 &6 & 7\\
		\hline\\[-1.5ex]
	Numeric & 4.451&4.509 &4.610 &4.683&NA & NA& NA \\[0.5ex]
	Eq.~(\ref{eq.offwb}) & 4.515 &4.515&4.565 &4.587&NA&NA&NA\\[0.5ex]
	\hline\hline
\end{longtable}

\vspace{-0.2in}
\subsection{The entangled mode}\label{gmode}
The PF gate is switched on the entangled mode in two steps, in the first step coupler frequency is brought out of $\wi$ value so that static $ZZ$ becomes nonzero. In the second step, qubits are driven externally by microwave pulse with amplitude $\Omega$. 

One two-qubit circuit example is to consider that we use microwave pulse to drive the `control' qubit, i.e. the qubit whose second excited level is higher, with the frequency of the other qubit, namely the `target' qubit. As discussed above, this driving is called cross resonance drive and is supposed to supply only the conditional $ZX$-type interaction between qubits,  however, the desired external force is accompanied by 3 types of unwanted operators  listed below:
\begin{enumerate}
	\item Classical crosstalk: supposedly triggered by the classical translation of driving electromagnetic waves to the position of target qubit,
	\item Control $Z$ rotation: due to driving control qubit with  qubits detuning frequency,  
	\item Microwave-assisted $ZZ$ interaction: triggered by $\Omega$-transition between computational and noncomputational levels ---listed in Eq. (\ref{eq. Horig}) and shown in  Fig.~\ref{fig:diag} --- that changes $\zeta$ level repulsion.  
\end{enumerate}

The first two can be eliminated as described in Ref. \cite{mckay2017efficient}: by driving target qubit with a second pulse to eliminate classical crosstalk, and either echoing the pulses or software-counterrotating control qubit to eliminate its detuning $Z$ rotation. However, these methods do not eliminate the third one. To eliminate it we proposed a method called ``dynamic freedom'',  which sets total $ZZ$ to zero by fine-tuning microwave parameters so that it cancels out the static parasitic interaction~\cite{xu2021zz-freedom}. The PF gate takes advantage of the dynamic freedom in the PF/E mode by combining microwave driving with a tunable coupler.

Let us recall that after eliminating the classical crosstalk and control $Z$ rotation,  external driving activates the Hamiltonian (\ref{eq. Horig}) with transitions within and outside of computational levels shown in Fig.~\ref{fig:diag}.  By block-diagonalizing the Hamiltonian to the computational subspace one can find the following simplified version:  
\beq
H_{d}(\Omega)=\alpha_{ZX} (\Omega) \hat{Z}\hat{X}/{2}+\zeta_d (\Omega) \hat{Z}\hat{Z}/{4}
\label{eq.zx}
\eeq

Perturbation theory  helps determine $\zeta_d$ and $\alpha_{ZX}$ in terms of driving amplitude $\Omega$. Results show that  $\alpha_{ZX}(\Omega)$ depends linearly on $\Omega$ in the leading order and $\zeta_d(\Omega)$ depends on $\Omega^2$ (For details see Eq.~(15) and Fig.~5,6 in \cite{xu2021zz-freedom}).    One may expect that higher order corrections can be worked out by adding terms that have larger natural number exponent, however comparing results with experiment has shown in the past that perturbation theory is not accurate beyond leading order \cite{magesan2020effective}. Alternatively we use a nonperturbative approach, the so-called Least Action (LA)~\cite{cederbaum1989block,magesan2020effective}.

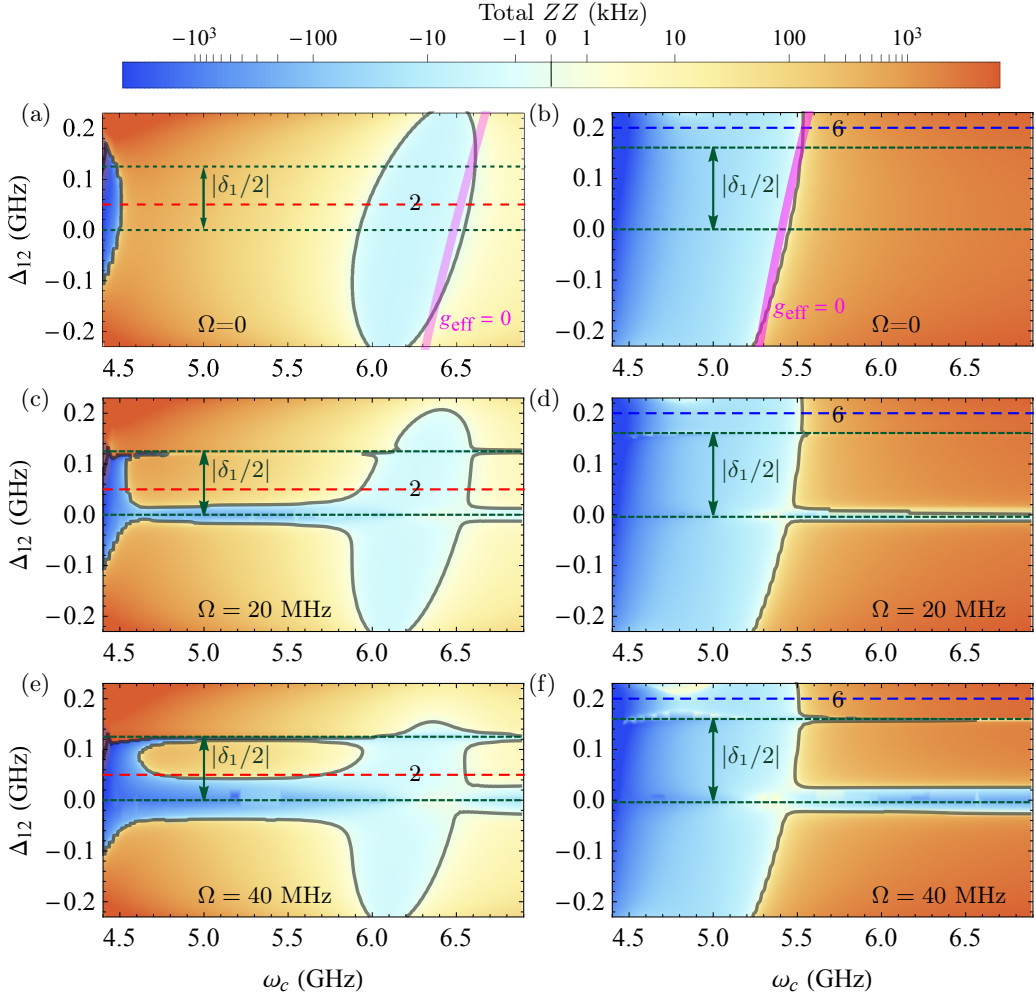
\begin{figure*}[t]
\hspace{0.3in}\includegraphics[width=0.7\textwidth]{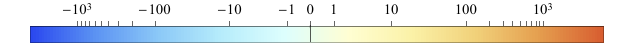}\put(-205,30){Total $ZZ$  (kHz)}\\
\includegraphics[width=0.4\textwidth]{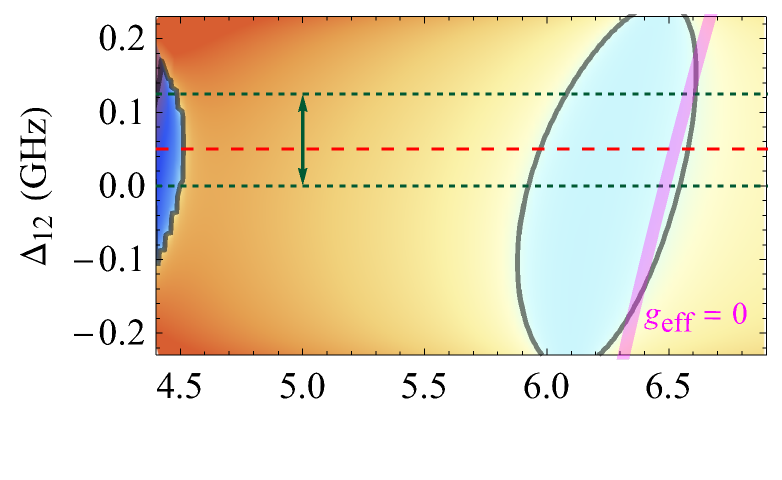}\hspace{-0.2in}
\includegraphics[width=0.4\textwidth]{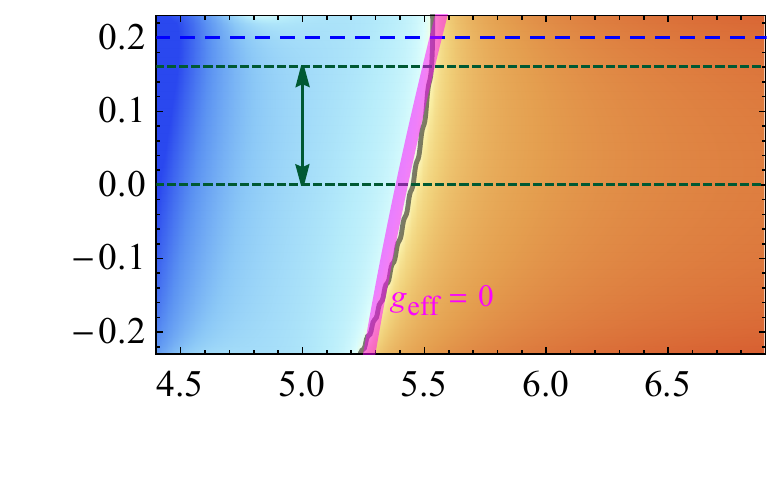}
\put(-387,118){(a)}\put(-195,118){(b)}
\put(-240,84){2}\put(-80,112){6}
\put(-320,38){$\Omega$=0}\put(-65,38){$\Omega$=0}
\put(-315,91){\textcolor{phthalogreen}{$|\delta_1/2|$}}\put(-122,91){\textcolor{phthalogreen}{$|\delta_1/2|$}}\\
\vspace{-0.25in}
\includegraphics[width=0.4\textwidth]{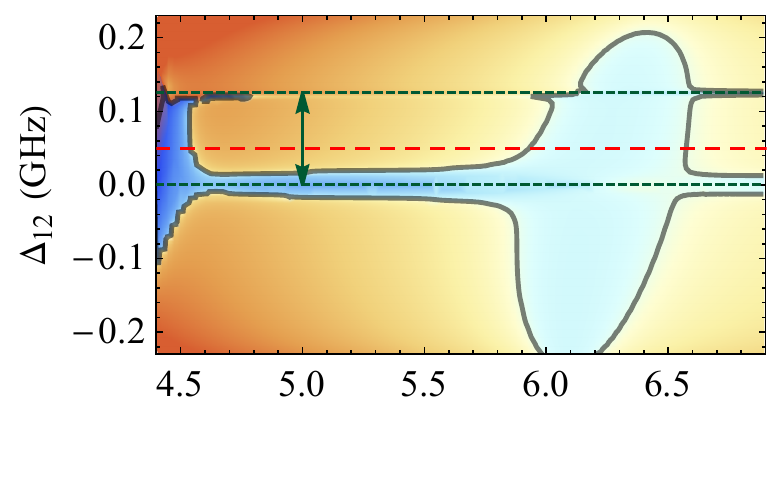}\hspace{-0.2in}
\includegraphics[width=0.4\textwidth]{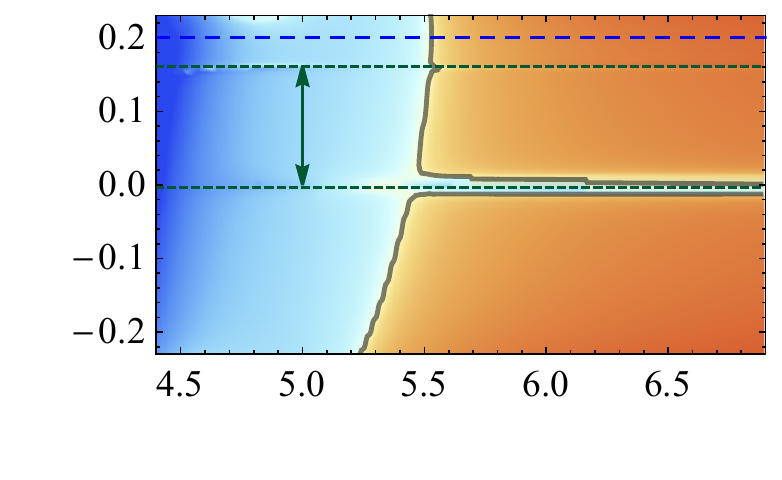}
\put(-387,118){(c)}\put(-195,118){(d)}
\put(-240,84){2}\put(-80,112){6}
\put(-320,38){$\Omega=20$ MHz}\put(-65,38){$\Omega=20$ MHz}
\put(-315,91){\textcolor{phthalogreen}{$|\delta_1/2|$}}\put(-122,91){\textcolor{phthalogreen}{$|\delta_1/2|$}}\\
\vspace{-0.25in}
\includegraphics[width=0.4\textwidth]{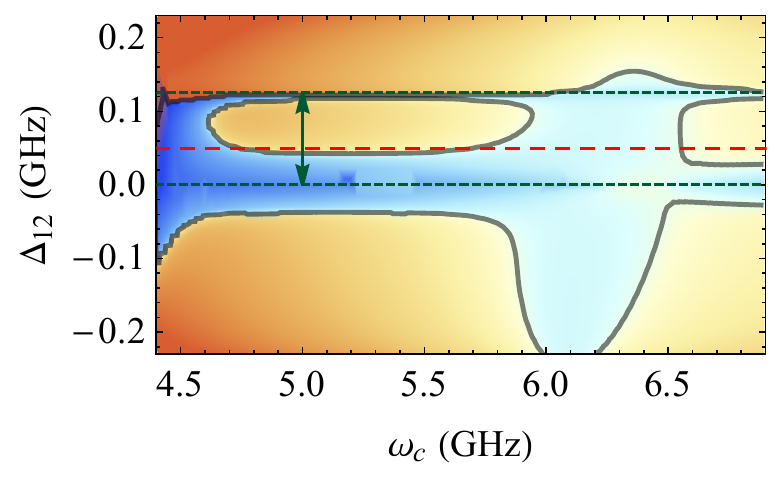}\hspace{-0.2in}
\includegraphics[width=0.4\textwidth]{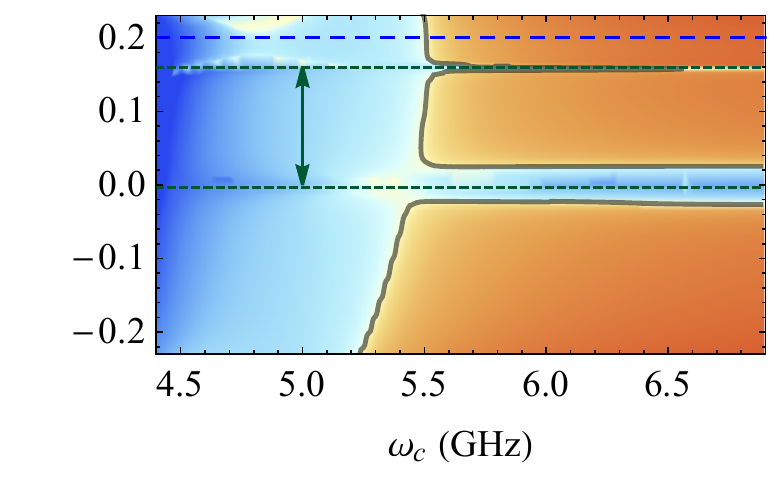}
\put(-387,118){(e)}\put(-195,118){(f)}
\put(-240,84){2}\put(-80,112){6}
\put(-320,38){$\Omega=40$ MHz}\put(-65,38){$\Omega=40$ MHz}
\put(-315,91){\textcolor{phthalogreen}{$|\delta_1/2|$}}\put(-122,91){\textcolor{phthalogreen}{$|\delta_1/2|$}}
  \vspace{-0.13in}
  \caption{Total $ZZ$ interaction as a function of qubits detuning frequency $\Delta_{12}$ and coupler frequency $\omega_c$ with parameters similar to device 2 in (a,c,e) and similar to device 6 in (b,d,f).  $\Omega=0$ in (a,b), $\Omega=20$~MHz in (b,c), and $\Omega=40$~MHz in (e,f). Red lines denotes the labelled devices 2 and blue lines denotes the labelled devices 6. Black boundaries are $ZZ$-free zones and magenta boundaries are $g_{\rm eff}$-free zones.\label{fig:dy2d}}
\end{figure*}

  Our numerical analysis evaluates total parasitic interaction $\zeta$ by adding the driving part to the static part.  We plot the total $ZZ$ interaction in Fig.~\ref{fig:dy2d} in a large range of qubits frequency detuning $\Delta_{12}$ and coupler frequency $\omega_c$ for two sets of circuit parameters. Left (Right) column plots show simulations for a set of parameters similar to device 2 (6) except that here we keep $\Delta_{12}$ variable. On the dashed lines labelled by 2 and  6 the detuning frequencies are fixed to values given in Table~\ref{tab:device}.  We plotted three sets of driving amplitudes in each row: Fig.~\ref{fig:dy2d}(a,b) show no driving $\Omega=0$ to study static level repulsions, Fig. \ref{fig:dy2d}(c,d) shows total $ZZ$ interaction after we apply driving with amplitude $\Omega=20$~MHz, and Fig.~\ref{fig:dy2d}(e,f) doubles the amplitude to $\Omega=40$~MHz.

 In these plots, we show the total parasitic interactions can be either positive (in red), or negative (in blue). The zero $ZZ$ devices are shown in black boundaries between the two regions.  In Fig. (\ref{fig:dy2d}) by closely examining $\zeta$ variation with $\omega_c$ one or more than one zero  points can be found for a device with fixed $\Delta_{12}$. More details with stronger driving amplitude can be found in Appendix~\ref{sec:dzzf}.
 
 In general, there are two types of $ZZ$-free  boundaries: Type I can be found in regions where $\zeta$ values become shallow by gradually being suppressed and they change the sign smoothly in white areas. Examples are zeros on the closed-loop in (a,c,e) and on the boundary in the middle of (b,d,f). In type II the $\zeta$ values abruptly change the sign between dark blue and red areas in a narrow domain of parameters. Examples are the far left side boundaries in (a,c,e). These correspond to two types of PF gate: {\it Genuine PF gate} starting from genuine idle mode to E mode with type I freedom and {\it Affine PF gate} starting from affine idle mode to E mode with type II freedom.

 %Perturbation theory does not find type II zeros because in their vicinity $\zeta$ function diverges and without passing through zeros they change signs. However detailed numerical analysis as shows that $\zeta$ is finite everywhere, which can also be understood from defining $ZZ$ by differences between energy eigenvalues, see below Eq.~(\ref{eq. Heff}).
 
% A subset of zeros is interaction-free by satisfying $g_{\rm eff}=0$.  We show the area where the effective interaction is zero by solid magenta line in Fig.~\ref{fig:dy2d}(a,b). Note that in the figures the scale of $ZZ$ values is kilohertz and usually $g_{\rm eff}$s are in MHz scales.  There are regions where the black and magenta boundaries overlap, at these parameters the two-qubit circuit is both  $ZZ$- and $g_{\rm eff}$-free and we reserve them as suitable candidates to be used as PF/I operational mode. 

 External driving in Fig.~\ref{fig:dy2d}(c-f) leaves a large class of devices with zero total $ZZ$ interaction, however by comparing total $ZZ$ with the static one it is noticed that external driving distorts the freedom boundaries.  In (a,c,e) subplots, which describe the same devices,  by increasing $\Omega$ the closed-loop surrounding a blue island on the right is shrunk, while a new closed-loop appears in the middle surrounding a red island. These boundaries are additionally distorted for devices with resonant frequency qubits $\Delta_{12}=0$ and devices at the symmetric point with  $\Delta_{12}=-\delta_1/2$. Perturbation theory shows that $\zeta$ diverges at these two points.  We show these points in darker green dashed lines. Our nonperturbative numerical results based on LA method shows $\zeta$ stays finite, however by increasing driving power, near these detuning values microwave-assisted component $\zeta_d$ is largely magnified and heavily dominates total $ZZ$ therefore zero boundaries are largely distorted.  Further discussion about the derivation of microwave-assisted components $\zeta_d$ can be found in Appendix~\ref{app.d26}.

\begin{figure}[tp]
	\centering
	\includegraphics[width=0.48\textwidth]{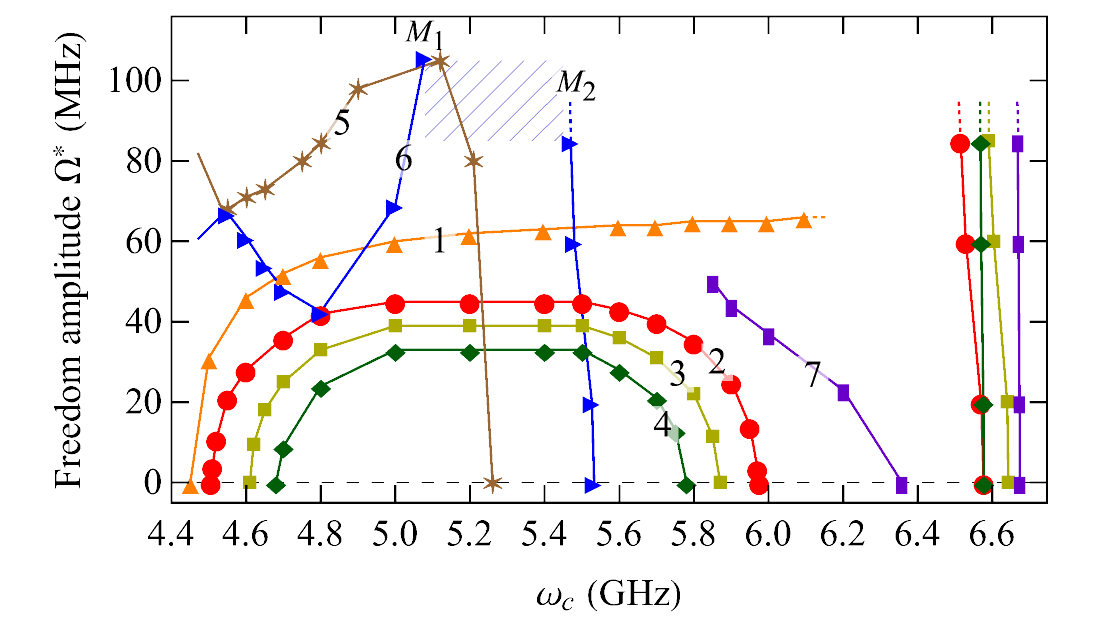}
	\put(-250,135){(a)}\\
	\vspace{-0.1in}
	\includegraphics[width=0.48\textwidth]{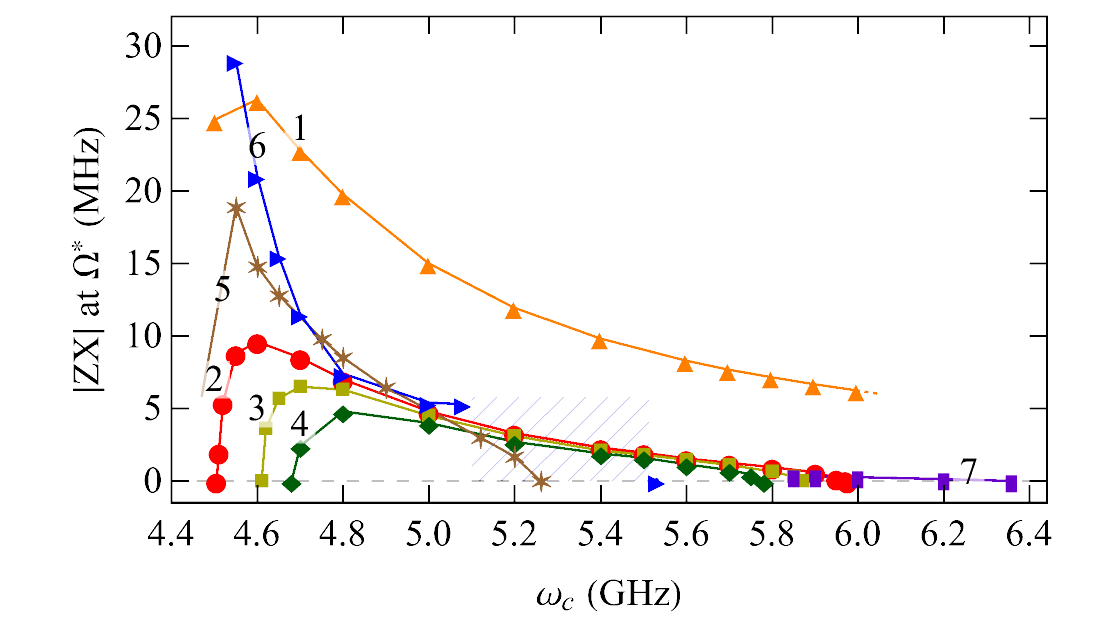}
	\put(-250,135){(b)}
	\vspace{-0.1in}
	\caption{(a) Freedom amplitude $\Omega^*$ as a function of the coupler frequency in devices 1-7.  (b) Corresponding $ZX$ rate. Shaded area between $M_1$ and $M_2$ indicates the absence of dynamic $ZZ$ freedom.}
	\label{fig:coupler2dy}
\end{figure}

Let us study the devices listed in Table \ref{tab:device} and we externally drive each with driving amplitude $\Omega$ and then evaluate the coupler frequency for dynamic freedom. For any amplitude associated to a $ZZ$-free coupler frequency is named the {\it{freedom amplitude}} denoted by $\Omega^*$. Figure~\ref{fig:coupler2dy}(a) shows driving device 1 with amplitudes below 60~MHz sets total parasitic interaction to zero. Devices 2-4 show three such frequencies in a rather weaker domain of freedom amplitudes, with an interesting feature on the rightmost one, near $\sim$6.6~GHz. Increasing driving amplitudes does not change the strongest $\omega_c^{\rm E}$, therefore at this frequency not only static level repulsion $\zeta_s$ is zero, but also driving assisted component $\zeta_d$ vanishes since $g_{\rm eff}=0$. Devices 5 and 6 show decoupling frequency below a certain driving amplitude, above which more $ZZ$-free coupler frequencies are added up. In device 6, however there is a frequency domain between $M_1$ and $M_2$ in which parasitic freedom is not expected to take place and we indicate it with the shaded region and we will come back to it later.  For device 7 staying beyond straddling regime, the static $ZZ$ freedom is only realized at higher coupler frequency.

A two-qubit gate not only is needed to have high fidelity, but also it must be fast because such gates can perform many operations during qubit coherence times. as discussed above it is important that external driving supplies a strong $ZX$ interaction, mainly because the strength of $ZX$ interaction, i.e. $\alpha_{ZX}$ scales inversely with the time that consumes to perform the gate, the so-called gate length $\tau \sim 1/\alpha_{ZX}$. Therefore the entangled mode of the PF gate must be tuned on a coupler frequency that not only is located on $ZZ$-free boundary but also present a short gate length.  Figure~\ref{fig:coupler2dy}(b) plots $ZX$ strength at all freedom amplitudes and shows that  $ZX$ rate is stronger at lower $\omega_c$'s. Therefore a $ZZ$-free coupler frequency with strong $ZX$ strength can be used for $\omega_c^{\rm PF/E}$. One exception is device 7 which stays out of straddling regime and has weaker $ZX$ rate, so it is not feasible to implement a CR-like gate and will not be discussed later.

%This result is interesting as it shows there is a sufficiently large separation between the coupler frequency of PF/E and PF/I modes.  Let us remind that the PF/I modes can be found on the right domain of $\omega_c$-axis where decoupling points are.

One of the advantages of our numerical analysis is that it can predict nonlinear correction in both $\zeta$ and $\alpha_{ZX}$ denoted in Eq.~(\ref{eq.zx}).   Perturbation theory determines the leading order of $\alpha_{ZX}$ and $\zeta_d$ are linear and quadratic in $\Omega$, respectively~\cite{xu2021zz-freedom}. The perturbative theory considers higher-order terms with next natural-number exponents above the leading terms, however in comparison with experiment those results are not accurate beyond leading order~\cite{magesan2020effective}. Since our approach is different we consider higher-order corrections in real-number exponents: 
\beqr     
\zeta_d (\Omega)&=&\eta_2\Omega^2+\eta_a\Omega^a, \  {\rm with} \ \ a>2,\\  
\alpha_{\rm zx} (\Omega) &=&\mu_1\Omega+\mu_b\Omega^b,\ \  {\rm with }\ \ b>1. 
\label{eq.zz}
\eeqr

\begin{figure}[h!]
	\centering
	\includegraphics[width=0.48\textwidth]{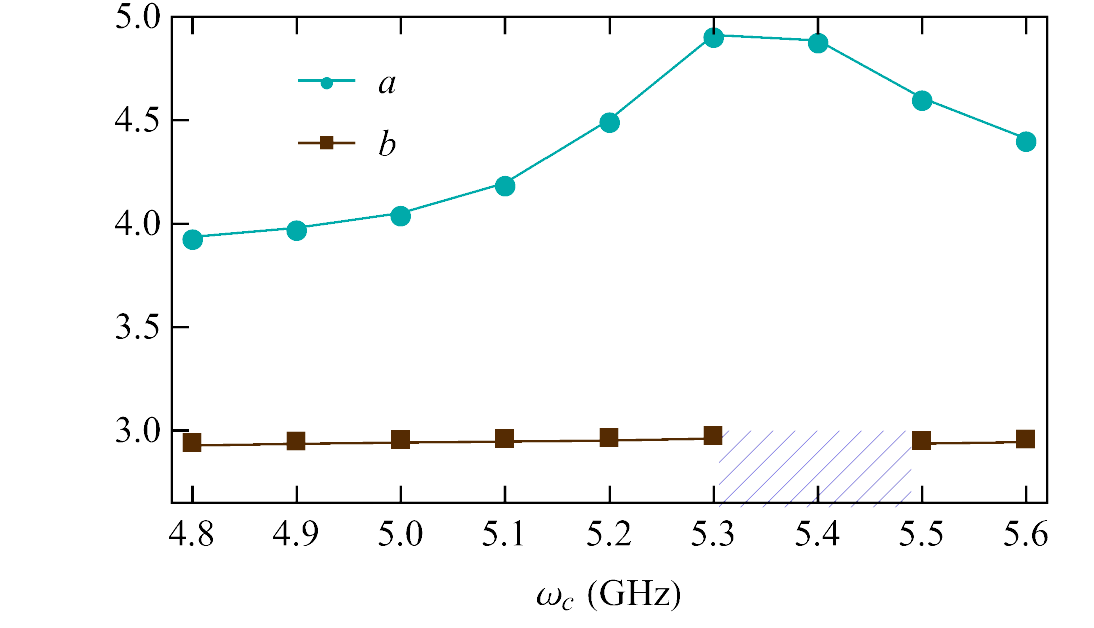}
	\vspace{-0.3in}
	\caption{Beyond 2nd (1st) order exponent of $\Omega^a$  ($\Omega^b$) in $\zeta_d$ ($\alpha_{ZX}$)  at different coupler frequency $\omega_c$ for device 6. Shaded area denotes the region where effective coupling $g_{\rm eff}$ is small and starts to change its sign.}
	\label{fig:horder}
\end{figure}

Our numerical results for $\zeta_d$ in device 6 estimates the exponents $a$ and $b$ at different coupler frequencies $\omega_c$.  The result is summarized in Fig.~\ref{fig:horder}, in which far-left points are similar to perturbative results; i.e. $a=4$ in $\zeta$ and $b\sim 3$ in $\alpha_{ZX}$.   However, there is a domain of frequency in which $a$ increases and nearly reaches 5. Moreover, within the same domain $ZX$ rate vanishes and this makes it meaningless to calculate $b$ exponent in that region. 

In device 6 the higher-order term of $\zeta_d$ has the opposite sign of $\eta_2 \Omega^2 +\zeta_s$ and is the dominant term in the coupler frequency domain 5.1--5.5~GHz. Therefore in this domain total $ZZ$ cannot vanish.  This describes the reason behind why there is a shaded area in Fig.~\ref{fig:coupler2dy}(a) in device 6 where PF/E mode cannot be found. More details can be found in Appendix~\ref{app.higherorder}.

\section{Error mitigation of PF gate}\label{sec:error}

Switching from Idle mode to entangled mode takes place in two steps: first coupler frequency is changed,  then microwave pulse is activated. The other way around needs to take place in the reversed order: first, the microwave pulse is switched off, then the coupler frequency is changed. In each step, there is a possibility that quantum states accumulate error.  Although PF gate can effectively eliminate universal $ZZ$ interaction, however, it still suffers from unwanted transitions during coupler frequency change.  Moreover, limited qubit coherence time can be another source of fidelity loss. Here we quantify the performance of both Genuine PF gate and Affine PF gate by calculating the metric of gate fidelity. 

\vspace{-0.2in}
\subsection{Error during coupler frequency variation}
   
\begin{figure}[tp]
	\centering
	\includegraphics[width=0.45\textwidth]{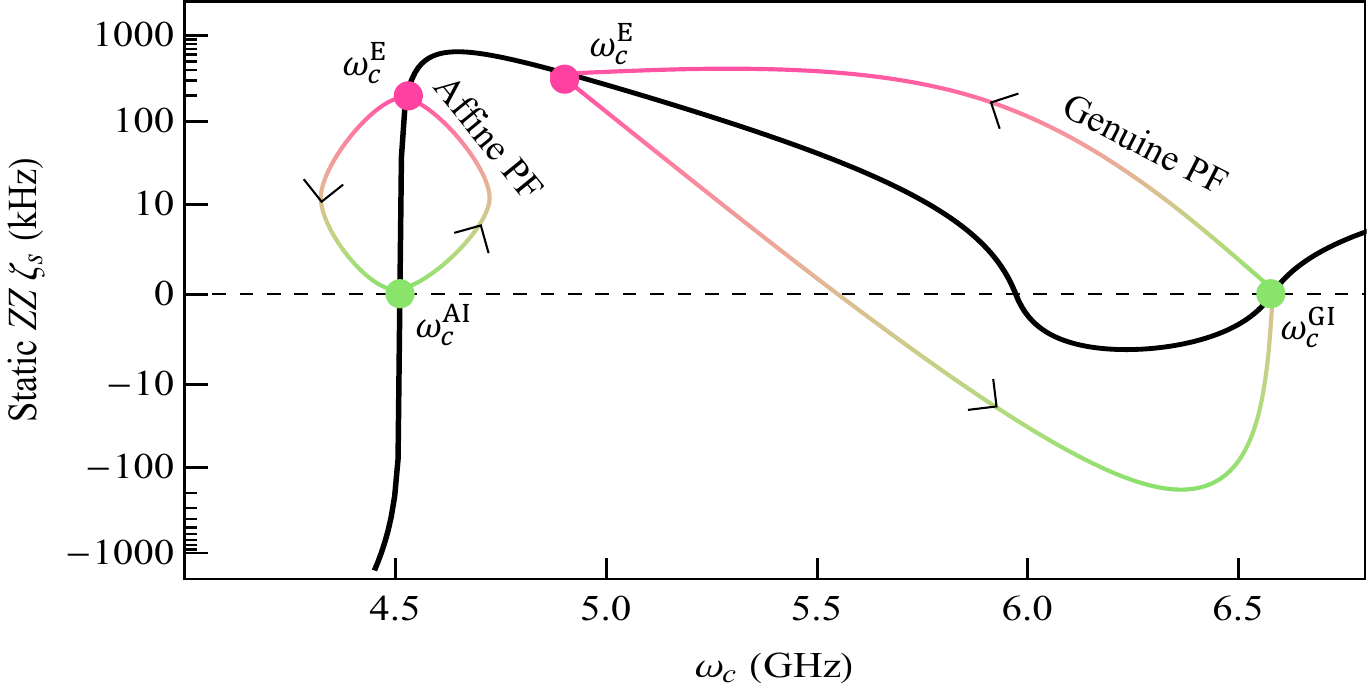}\put(-244,115){(a)}\\
	\vspace{0.02in}
	\hspace{0.05in}\includegraphics[width=0.23\textwidth]{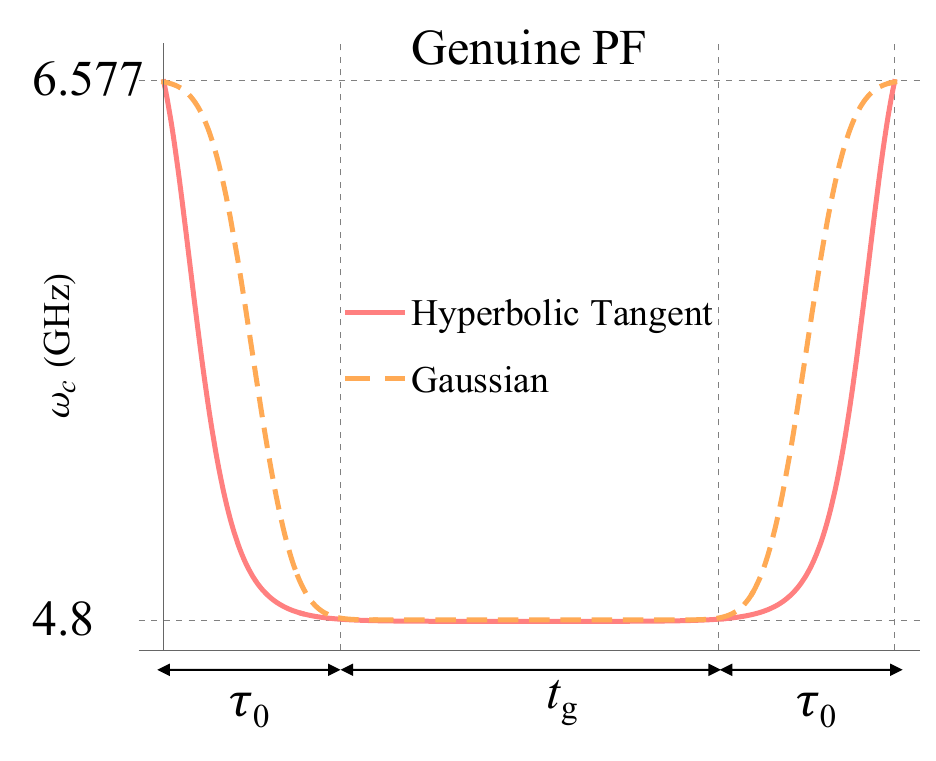}\put(-128,85){(b)}\hspace{0.05in}
	\includegraphics[width=0.23\textwidth]{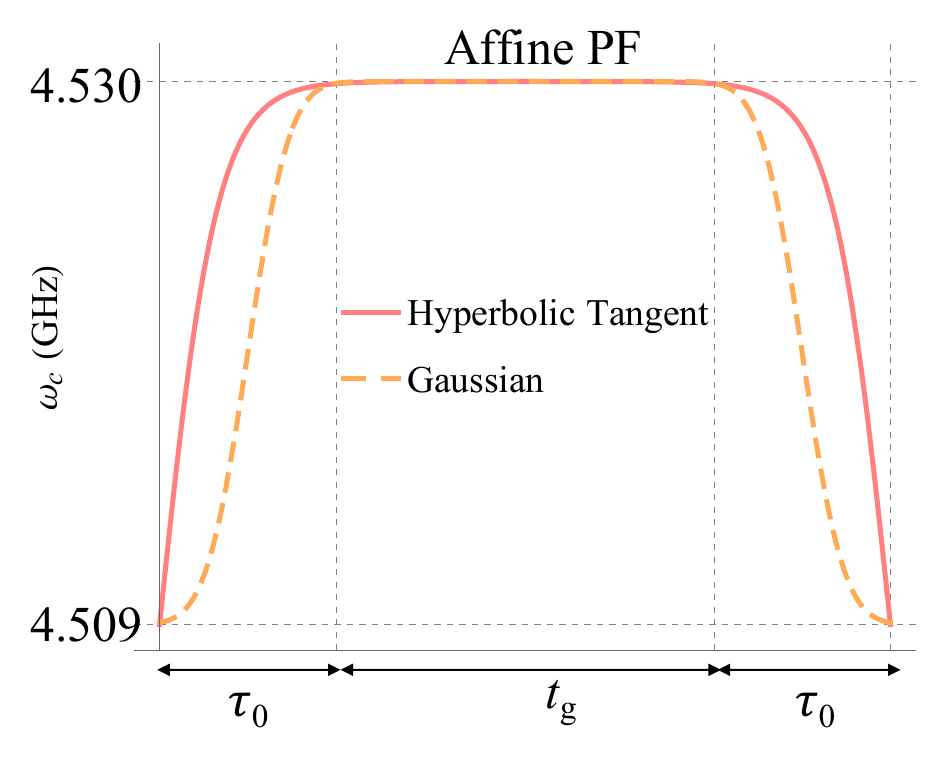}\put(-125,85){(c)}\\
	\includegraphics[width=0.24\textwidth]{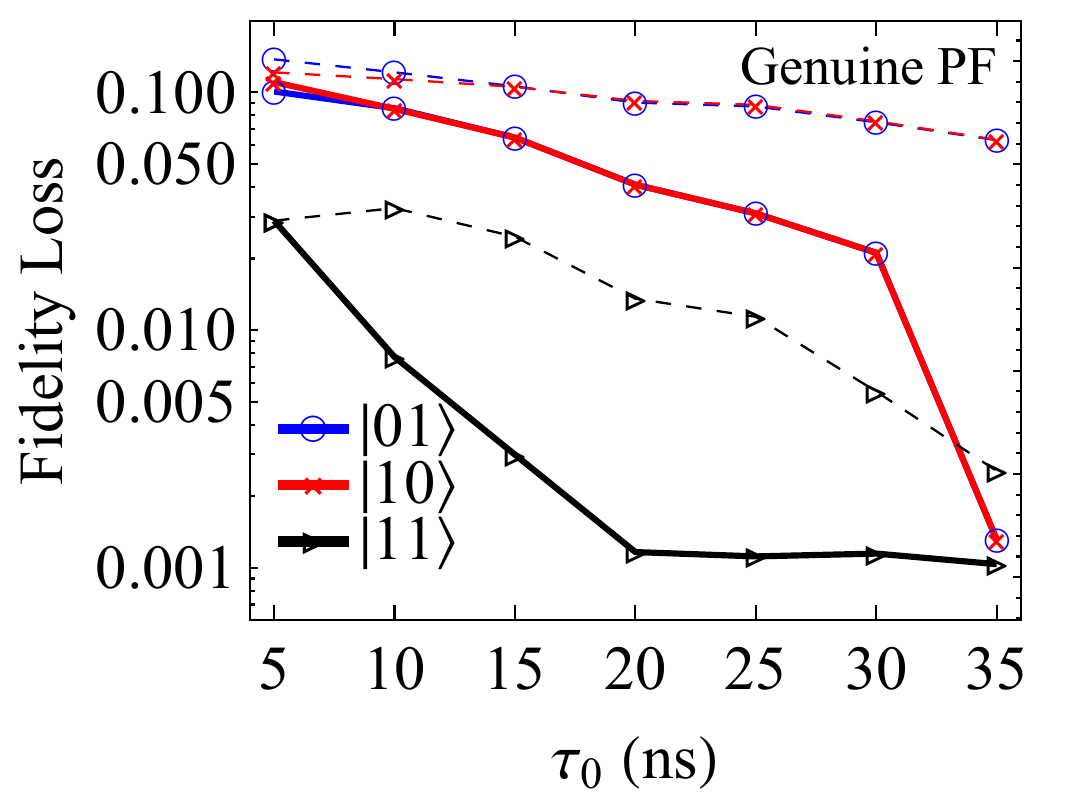}\put(-128,90){(d)}
	\includegraphics[width=0.24\textwidth]{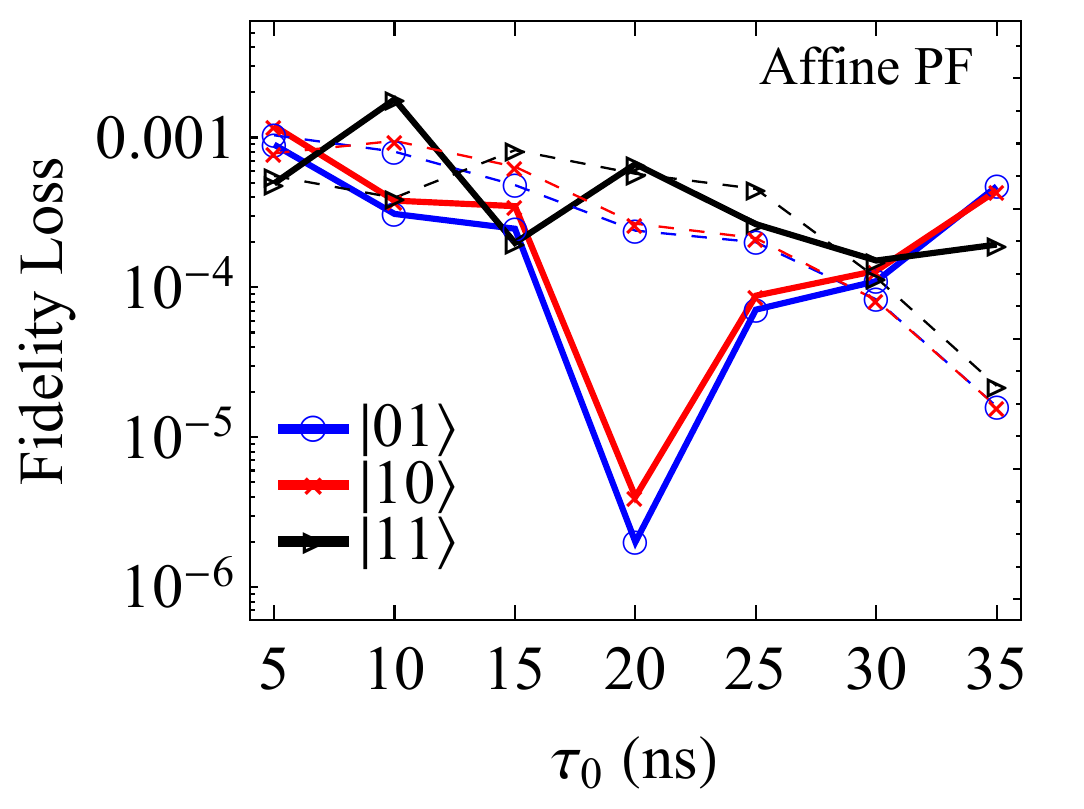}\put(-125,90){(e)}
	\vspace{-0.1in}
	\caption{(a) Sketches of two types of PF gate: start from idle mode, then switch to the entangled mode and finally go back to the idle mode.
	(b) Coupler frequency switching protocols for Genuine PF gate with the following two pulse envelopes: Hyperbolic tangent envelope (solid) and Flat-topped Gaussian envelope (dashed). (c) Coupler frequency switching protocols for Affine PF gate with similar pulse envelopes. (d) Fidelity loss of the computational states for Genuine PF gate due to leakage from the two pulse shapes. (e) Fidelity loss of the computational states for Affine PF gate due to leakage from the two pulse shapes. \label{fig:adia}}
\end{figure}

For the Genuine PF gate, the coupler frequency is far detuned from the frequency of qubits. Switching the mode to entangled mode requires that the coupler frequency is brought much closer to the qubits. While for the Affine PF gate, the coupler frequency is near qubits, then switching the mode to entangled mode only needs slight change in the coupler frequency, e.g. with a WTQ.  Figure~\ref{fig:adia}(a) sketches the two types of PF gate implemented in device 2. In either case to avoid reinitialization of qubit states after a frequency change, we can perform the frequency change so that leakage does not take place from the computational subspace to other energy levels. This mandates to perform the coupler frequency change adiabatically \cite{dicarlo2009demonstration}. 

Leakage rate out of computational levels depends on the ramping speed of coupler frequency $d\omega_c/dt$. In particular, if the coupler frequency is tuned by external magnetic flux $f=\Phi_{\rm ext}/\Phi_0$ with $\Phi_0$ being flux quantum unit, the rate of coupler frequency change can be written in terms of $d f/dt$ \cite{xu2020high-fidelity}. Here we compare two protocols for the pulse envelopes to quantify the leakage due to $\omega_c$ modulation. 

\paragraph*{Genuine PF gate:} Figure~\ref{fig:adia}(b) shows two pulses that we prepared for being used on device 2: a hyperbolic tangent envelope pulse in solid line and a flat-top Gaussian envelope in dashed line.  The qubits are decoupled at $\wia=$6.577~GHz. Figure~\ref{fig:coupler2dy}(b) shows that $\alpha_{ZX}$ is rather strong --- nearly $\sim$5~MHz --- at the frequency 4.8~GHz which we take as $\wE$.  Note that much stronger $\alpha_{ZX}$ of about nearly 10~MHz is also possible on this device which corresponds to 0.2~GHz smaller coupler frequency. 

\paragraph*{Affine PF gate:} Figure~\ref{fig:adia}(c) shows similar two types of pulse envelope. The difference is that for Affine PF gate idle coupler frequency $\wib$ is lower than $\omega_c^{\rm E}$. On device 2 qubits are $ZZ$-free at $\wib=$4.509~GHz. To make a comparison, we tune the coupler frequency to make $\alpha_{ZX}$ also around 5~MHz but much closer to $\wib$ --- at the frequency 4.530~GHz which we take as $\wE$.  

Let us denote the total time it takes for the PF gate to start at  $\wi$ and return to it by  $t=2\tau_0+t_g$ in which $t_g$ is the microwave activation time in between two coupler frequency changes. Each coupler frequency change takes place during time $\tau_0$. By varying $\tau_0$ and solving differential equations in the open quantum system, we can evaluate fidelity loss for computational states and then determine the optimized pulse for frequency change. 

Here we calculate the fidelity loss of computational states in absence of driving pulses to evaluate the errors from the switching idle and entangled modes. Figure~\ref{fig:adia}(d) shows the fidelity loss of states $|01\rangle$, $|10\rangle$, $|11\rangle$ for Genuine PF gate by varying $\tau_0$ during the I-E-I journey without external drive. Both pulses show that overall by increasing $\tau_0$ the computational state fidelity increases. However there is a difference between the two pulse performances.  The hyperbolic tangent envelope pulse which rapidly changes coupler frequency between I and E modes can further reduce the error by raising all computational state fidelities to above $99.9\%$ at  $\tau_0=35$~ns. For Affine PF gate, the individual state fidelity loss is less than 0.1\% as shown in  Fig.~\ref{fig:adia}(e),  in particular, the shortest rise/fall time for achieving 99.9\% total state fidelity is $\tau_0=15$~ns for the hyperbolic tangent envelope .

%both envelopes can further reduce the error to around $99.99\%$ at  $\tau_0=30$~ns. 

\vspace{-0.2in}
\subsection{Gate error during  external driving}

\begin{figure}[bp]
	\centering
	\includegraphics[width=0.48\textwidth]{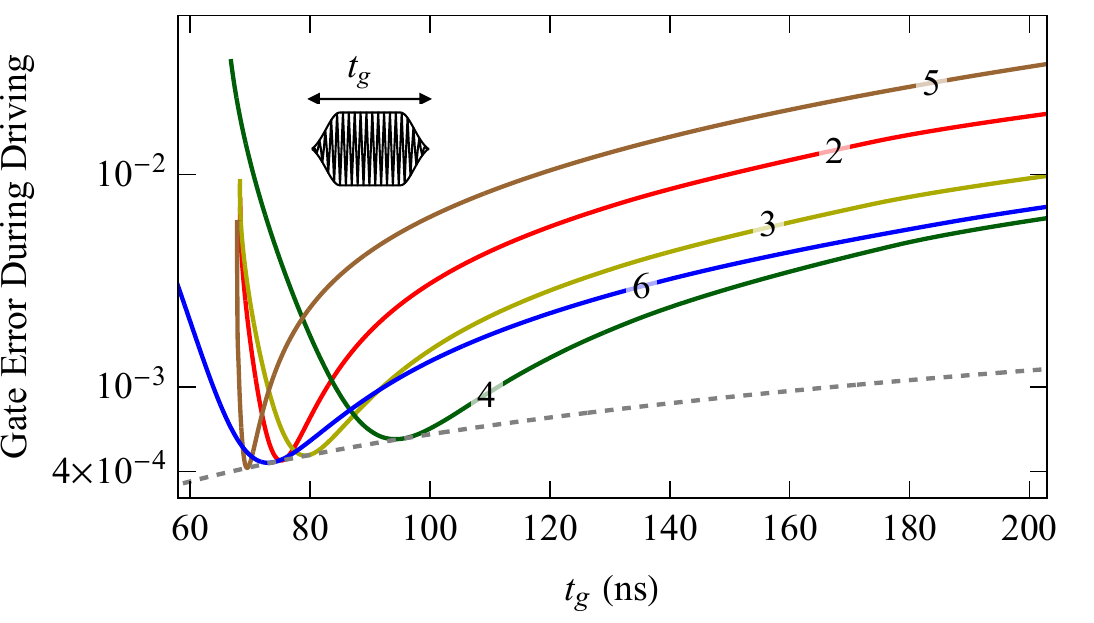}\put(-250,140){(a)}\put(-70,40){Genuine PF}\\
	\includegraphics[width=0.48\textwidth]{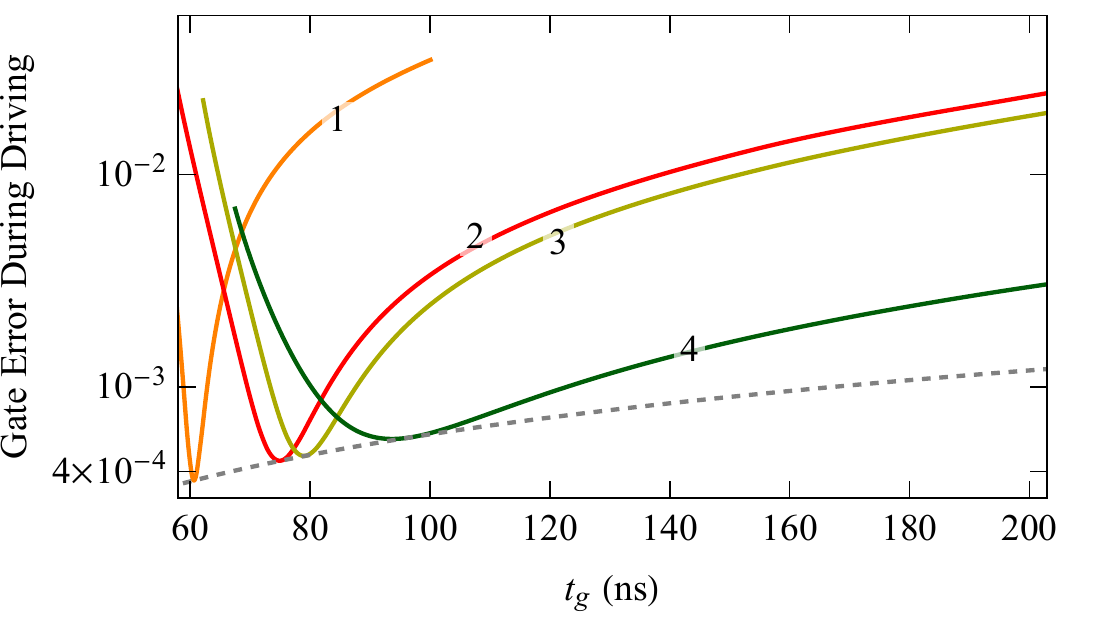}\put(-250,140){(b)}\put(-70,40){Affine PF}
	\vspace{-0.1in}
	\caption{(a) Genuine PF gate error during external driving versus gate length $t_g$. The coupler frequency is parked at $\omega_c/2\pi=4.8$ GHz. Inset plot is the round squared CR pulse shape with $\sim$20~ns rise and $\sim$20~ns fall times. (b) Affine PF gate error during external driving versus gate length $t_g$. The coupler frequency is parked at 4.472, 4.530, 4.658, 4.731~GHz for devices 1-4, respectively. \label{fig:error}}
\end{figure}

In the two pulses discussed in Fig. \ref{fig:coupler2dy}(a) once the coupler frequency is changed to a lower value the circuit is ready to experience a $ZZ$-free $ZX$-interaction. This takes place by turning on the microwave during time $t_g$. The length of $t_g$ for $ZX$ gate is governed by microwave driving. During the time $t_g$ the qubit pairs enjoy the absolute freedom from total $ZZ$ interaction, however, gate fidelity is limited by qubit coherence times. 

Let us indicate here that we do not consider the option of echoing the microwave driving as this doubles the gate length. Instead, we follow the recent practice at IBM where a single cross resonance driving is applied on qubits followed by a virtual $Z$ rotation~\cite{mckay2017efficient}. We also take the example of $ZX_{90}$ pulse for this typical analysis. In this pulse the relation between $\alpha_{ZX}$ and the length of the flat top in the microwave pulse $\tau$ has been discussed before Section (\ref{sec:realization}), i.e. $\tau=1/4\alpha_{ZX}$. Here we consider the microwave pulses are round squared with $\sim$20~ns rise and $\sim$20~ns fall times as shown in the inset of Fig.~\ref{fig:error}. Thus the total $ZX$-gate length is $t_g=(40+1/4\alpha_{ZX}[\rm MHz])$ in nanoseconds. 

Coherence times of Q1 and Q2 are ideally assumed to be all the same: i.e. $\{T_1^{(1)}, T_2^{(1)}\}=\{T_1^{(2)}, T_2^{(2)}\}=\{200,200\}$~ns. For Genuine PF gate, we consider the coupler frequency at E mode $\wE$ is 4.8~GHz. With these parameters, the $ZX$-gate error rates are plotted for devices 2-6 in Fig.~\ref{fig:error}(a). All error rates have a common behaviour as they show a minimum at a certain gate length $t_g$ where the error rate is as low as that expected only from coherence time limitation where the device experiences total $ZZ$ freedom. Among the five devices 2-6 plotted, device 5 (a hybrid CSFQ-transmon) has the shortest gate length and after that stands device 6 (a pair of transmons with 200~MHz frequency detuning). The minimum error rate, although cannot be eliminated without perfecting individual qubits, indicates the possibility of $ZX$ interaction gate with fidelity as high as $99.9\%$. Note that there is a limitation on the gate length behaving as a cutoff in Fig.~\ref{fig:error}, see Ref.~\cite{xu2021zz-freedom} for more details.

For Affine PF gate, we tune the coupler frequency such that $ZX$ rate in devices 2-4 is the same as that in PF-G gate at corresponding freedom amplitude, and plot the error rate in  Fig.~\ref{fig:error}(b). The difference compared to Genuine PF gate is the coupler is only tuned in a narrow domain e.g. 50 MHz using a WTQ, which can effectively suppress decoherence from flux noise. Moreover, required driving amplitude for the same gate duration is weaker, then total $ZZ$ is smaller and less detrimental to the gate fidelity. Here we also study device 1 with the same qubit detuning as device 2. Figure~\ref{fig:error}(b) shows that device 1 enables the Affine PF gate with stronger $ZX$ rate and therefore shorter gate length and less error rate.

Summing the error rate from both rise/fall times and decoherence, all together for device 2 one can estimate a minimum total time length of 105~ns long Affine PF gate that takes Q1 and Q2 from a $ZZ$-free affine idle mode to entangled mode and returns it back to original affine idle mode only by weakly tuning the coupler. The idle to idle error rate during the affine PF gate time of $t_g+2\tau_0$ will be about $0.1\%$. While for the Genuine PF gate the minimum gate length is 135 ns with 99.7\% gate fidelity.

There is a possibility that the length of the PF gate can become even shorter if the microwave rise and fall times combine the two switching coupler frequency times. For the example discussed above this can save up to 40~ns from the gate length. Theoretically, such a time saving needs careful analysis in optimal control theory, which goes beyond the scope of this paper, however experimentally it can be investigated.

\vspace{-0.2in}
\section{Summary}
To summarize, we propose a new two-qubit gate by combining idling and entangling gates. This gate can safely switch qubits states between idle and $ZX$-entangled modes and once at both modes quantum states do not accumulate conditional $ZZ$ phase error in time.  By zeroing $ZZ$ in both modes qubits before, during, and after entanglement are safely phase-locked to their state. This gate can be realized in superconducting circuits by combining tunable circuit parameters and external driving in two ways: 1) At genuine idle mode tuning circuit parameter makes qubits decoupled and therefore the static $ZZ$ interaction vanishes. At entangled mode, the static $ZZ$ interaction is cancelled by a microwave-assisted $ZZ$ component, so that qubits are left only with the operation of $ZX$-interaction; 2) At affine idle mode qubits are strongly coupled, but the level repulsions from both sides of computational space cancel each other. At entangled mode qubits $ZX$-interact with zero-$ZZ$. 

%Owing to the coupler tunability the strength of $ZX$ interaction can become stronger, interestingly, by modulating coupler frequency farther from idle mode. This makes the two modes to be separated by a high energy barrier and boost the state fidelity for a fast gate.

We evaluate a typical time length for the PF gate once its fidelity is only limited by qubit coherence times. In a complete operational cycle from idle to idle mode, in passing once through an entangled mode, the Affine PF gate is as short as 105~ns with the overall error budget rate being about 0.1$\%$. 

We show that this gate is universally applicable for all types of superconducting qubits, such as all transmon or hybrid circuits,  and certainly not limited to frequencies in the dispersive regime. We believe the PF gate will pave a new way to implement high-quality quantum computation in large-scale scalable quantum processors. 
\vspace{-0.2in}
\section*{Acknowledgement}
The authors thank Britton Plourde, Thomas Ohki, Guilhem Ribeill, Jaseung Ku, and Luke Govia for insightful discussions. We gratefully acknowledge funding by the German Federal Ministry 
of Education and Research within the funding program "Photonic Research
Germany" under contract number 13N14891, and within the funding program "Quantum Technologies - From Basic Research to the Market" (project GeQCoS), contract number
13N15685.

\appendix

\vspace{-0.2in}
\section{Comparing numerical and perturbation methods}\label{app compare}
The circuit Hamiltonian in the lab frame is written in the form of multilevel systems as 
\begin{eqnarray}\label{eq.hcoupler}
H_0=&&\sum_{i=1,2,c}\sum_n\omega_i(n_i)|n_i+1\rangle\langle n_i+1|+\sum_{i<j} \sum_n\nonumber\\
&&\sqrt{(n_i+1)(n_j+1)} g_{ij}\left(|n_i,n_j\rangle\langle n_{i}+1,n_{j}+1|\right.\nonumber\\
&&\left.-|n_{i}+1,n_j\rangle\langle n_i,n_{j}+1|+{ H.c.}\right),
\end{eqnarray}
where $\omega_i(n_i)=E_i(n_i+1)-E_i(n_i)$ and $\delta_i(n_i)=E_i(n_i+2)-2E_i(n_i+1)+E_i({n_i})$ with $E_i(n_i)$ being the bare energy of level $n$ for subsystem $i~(i=1,2,c)$. Especially, frequency and anharmonicity can be simplified as $\omega_i(0)=\omega_i$ and $\delta_i(0)=\delta_i$. 

\begin{figure}[h!]
	\centering
	\includegraphics[width=0.48\textwidth]{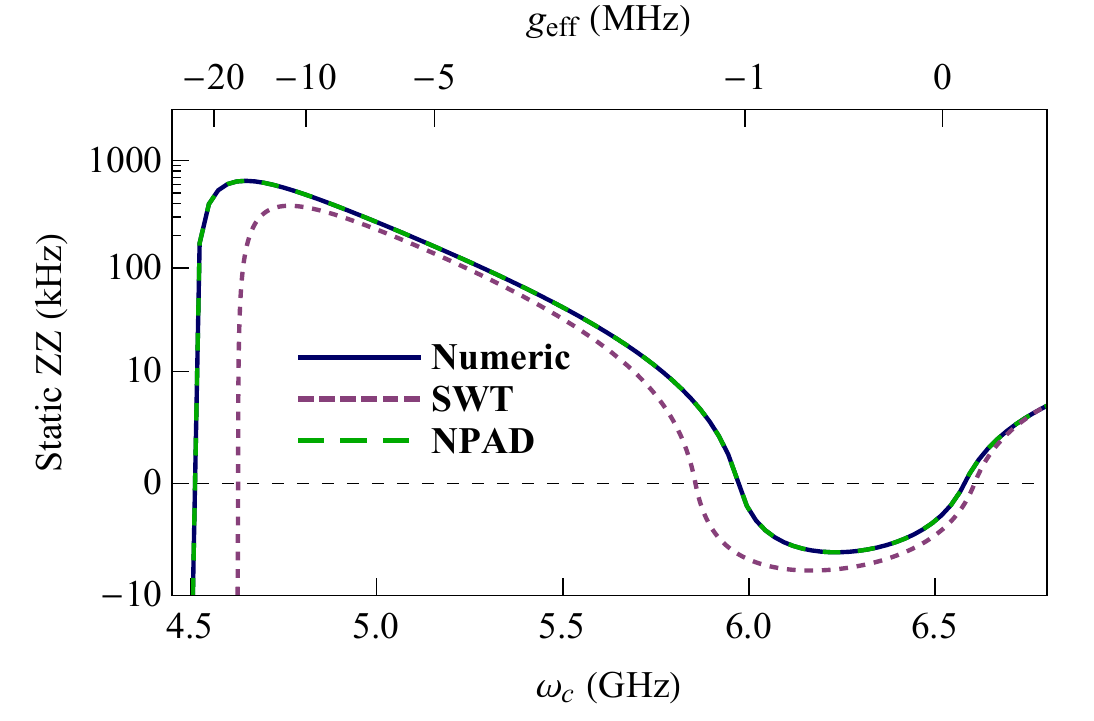}
  	\vspace{-0.3in}
	\caption{Static $ZZ$ interaction versus coupler frequency, SWT result is from Eq.~(\ref{eq.zeta}), NPAD makes use of the Jacobi iteration~\cite{li2021non-perturbative} and exact result is obtained by diagonalizing the Hamiltonian in Eq.~(\ref{eq.hcoupler}). The top axis is corresponding effective coupling $g_{\rm eff}$. The used circuit parameters are the same as device 2 with $\alpha_1=\alpha_2=0.022$.}
	\label{fig:staticzz}
\end{figure}

 We evaluate static $ZZ$ on the seven devices listed in Table~\ref{tab:device} by fully diagonalizing their corresponding circuit Hamiltonian and plotted results in Fig.~\ref{fig:coupler2zz}. Moreover, we compare static $ZZ$ interaction in device 2 with the following three methods: numeric simulation (Numeric), NPAD~\cite{li2021non-perturbative} and Schrieffer-Wolff Transformation (SWT). Figure~\ref{fig:staticzz} shows the static $ZZ$ at the lower $x$ axis for the coupler frequency, as well as  $g_{\rm eff}$ at the upper $x$ axis for the effective coupling strength.

\vspace{-0.2in}
\section{Driven Hamiltonian}\label{app.d26} 
When microwave drive is on, CR driving Hamiltonian $H_d=\Omega\cos(\tilde{\omega}_2 t)\left(|n_1\rangle\langle n_{1}+1|+|n_1+1\rangle\langle n_1|\right)$ needs to be transferred to the same regime as the the qubit Hamiltonian. In the rotating frame the total Hamiltonian is 
\begin{equation}
H_r=W^{\dagger}(\tilde{H_0}+\tilde{H_{ d}}) W-i W^{\dagger}\dot{W}.
\label{eq.Hr}
\end{equation}
where $\tilde{H_0}=U^{\dagger}H_0U$ with $U$ being the unitary operator that fully diagonalizes $H_0$, $\tilde{H_d}=U^{\dagger}H_dU$ and $W=\sum_{i=1,2,c}\sum_n \exp(-i\omega_d t \hat{n}_i)|n_i\rangle\langle n_i|$. 
For simplicity we assume $g_{1c}=g_{2c}=g$, $g_{12}=0$ and $\delta_1=\delta_2=\delta$, the transition rates in Eq.~(\ref{eq. Horig}) then are derived from perturbation theory and listed in Table~\ref{tab:lambda}.

\begin{longtable}{@{\extracolsep{\fill}}cc@{}}
	\caption{Transition rates}
	\label{tab:lambda}
	\endfirsthead
	\endhead
	\hline\hline\\[-2ex]
	\centering
	$\lambda_1$&$ -g^2 \delta  /  2\Delta_{12} \Delta_{2} (\Delta_{12} +\delta)$\\
	$\lambda_2$&$ -g  \delta  / 2\Delta_{1} (\Delta_{1}+\delta)$\\
	$\lambda_3$&$ -\sqrt{2} g^2  \delta  /\Delta_{12} (\Delta_{12}-\delta) (\Delta_{2}+\delta)$\\
	$\lambda_4$&$ -g   /2 \Delta_{1} $\\
	$\lambda_5$&$ -g^2  \delta  /\sqrt{2} \Delta_{12} \Delta_{2} (\Delta_{12}+\delta)$\\ 
	$\lambda_6$&$ g  \delta  / \sqrt{2} \Delta_{1} (\Delta_{1}+\delta)$\\
	$\lambda_7$&$ -g^2   (\Delta_{2}+\delta_c) /2 \Delta_{12} \Delta_{2} (\Delta_{2}-\delta_c)$\\ 
	$\lambda_8$&$-g   / \sqrt{2} (\Delta_{1}-\delta_c)$\\ 
	$\lambda_9$&$-g  \delta /\sqrt{2}\Delta_{1}(\Delta_{1}+\delta)$\\ 
	$\lambda_{10}$&$ g^2 /\Delta_{2}(\Delta_{12}+\delta)$\\ 
	$\lambda_{11}$&$-g^2  \delta/\Delta_{12} (\Delta_{12}-\delta)(\Delta_{2}+\delta)$\\ 
	\hline\hline
\end{longtable}

In the entangled mode only qubits are encoded, we can further simplify the total Hamiltonian by decoupling the tunable coupler,  block diagonlizing the Hamiltonian and then rewriting it in terms of Pauli matrices as discussed in Sec.~\ref{gmode}. 
\vspace{-0.2in}
\section{Dynamic $ZZ$ freedom}\label{sec:dzzf}
\begin{figure}[h!]
	\centering
	\includegraphics[width=0.48\textwidth]{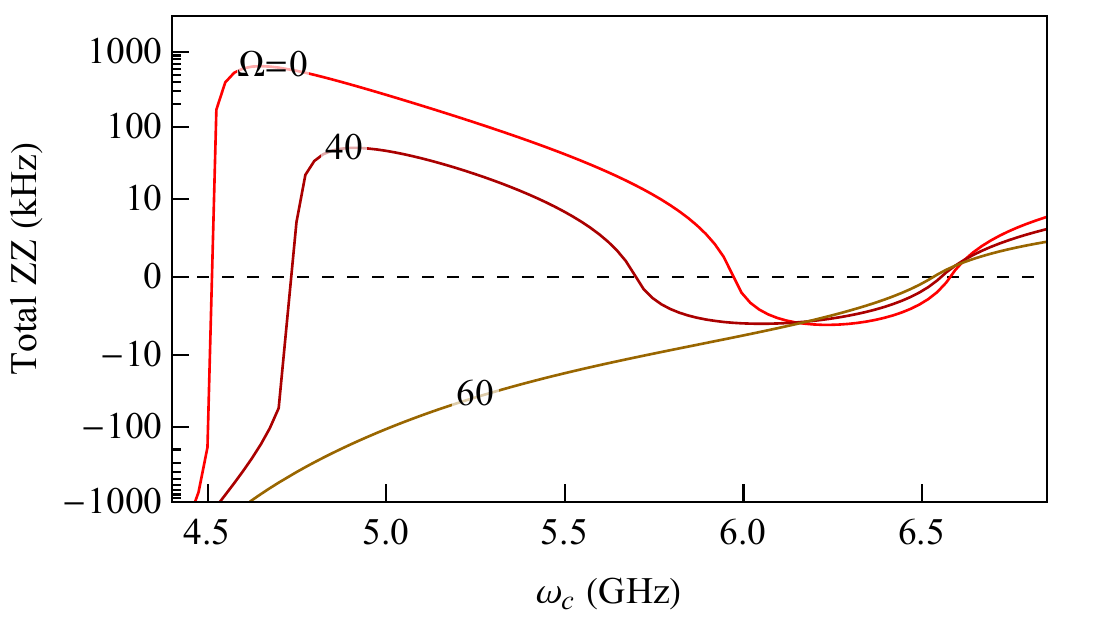}\put(-250,135){(a)}\put(-60,35){Device 2}\\
	\vspace{-0.05in}
	\includegraphics[width=0.48\textwidth]{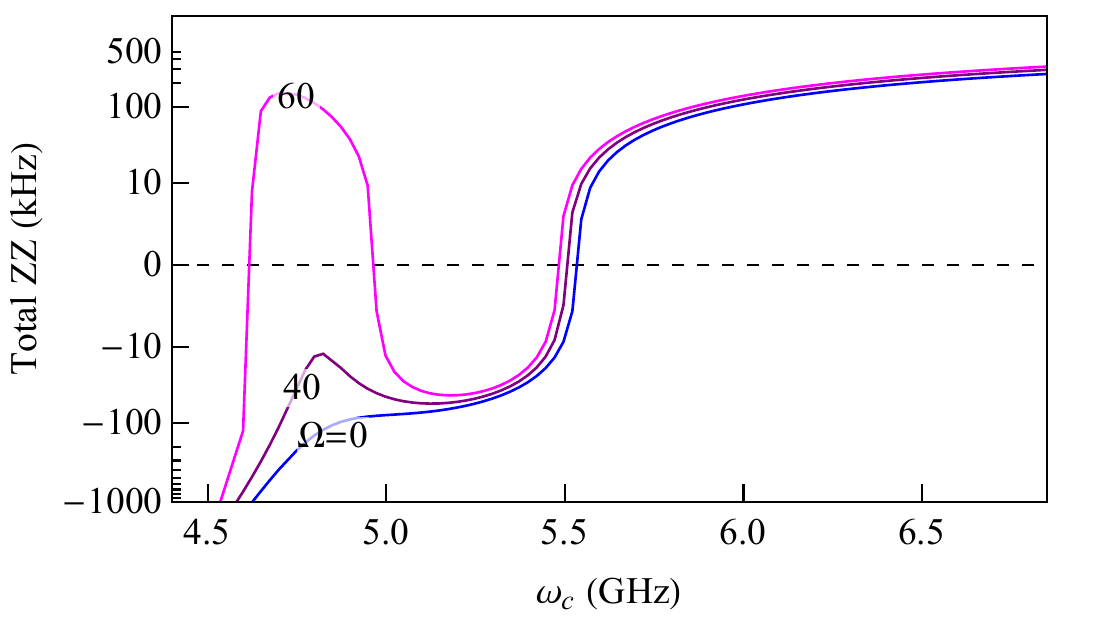}
	\put(-250,135){(b)}\put(-60,35){Device 6}
	\vspace{-0.2in}
	\caption{Total $ZZ$ interactions versus coupler frequency at different driving amplitude on (a) device 2 and (b) device 6.}
	\label{fig:dr26}
\end{figure}

Figure~\ref{fig:dr26} shows how the driving amplitude $\Omega$ impacts the total $ZZ$ interaction. In device 2, static $ZZ$ interaction exhibits three zero $ZZ$ points in terms of the coupler frequency. By increasing the driving amplitude, total $ZZ$ interaction becomes smaller and finally annihilates two of the zero $ZZ$ points, leaving the only one $\wia$. However, the behaviour of device 6 is opposite, external drive makes it possible to realize $ZZ$ freedom beyond the only $\wia$ point. 

\begin{figure*}
	\centering
	\includegraphics[width=0.7\textwidth]{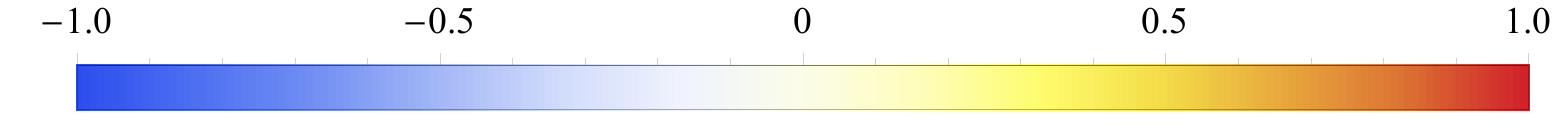}
	\put(-190,35){$\cos(\zeta \tau_p)$}\\
	\includegraphics[width=0.4\textwidth]{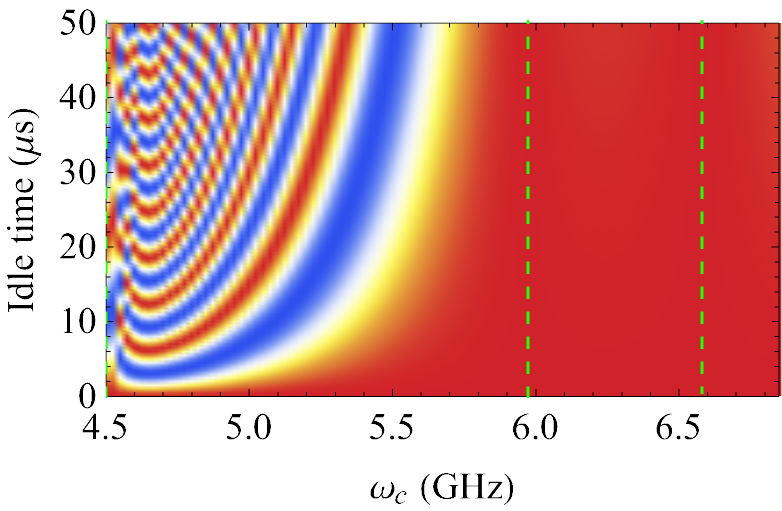}
	\includegraphics[width=0.4\textwidth]{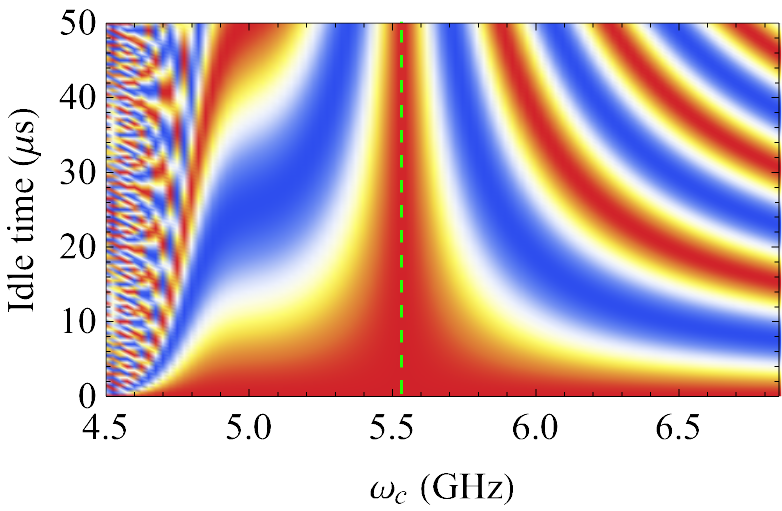}
	\put(-410,130){(a)}\put(-300,40){$\Omega=0$}\put(-245,35){\textcolor{green}{$g_{\rm eff}=0$}}
	\put(-205,130){(b)}\put(-140,40){$\Omega=0$}\put(-115,35){\textcolor{green}{$g_{\rm eff}=0$}}\vspace{-0.3in}\\
	\includegraphics[width=0.4\textwidth]{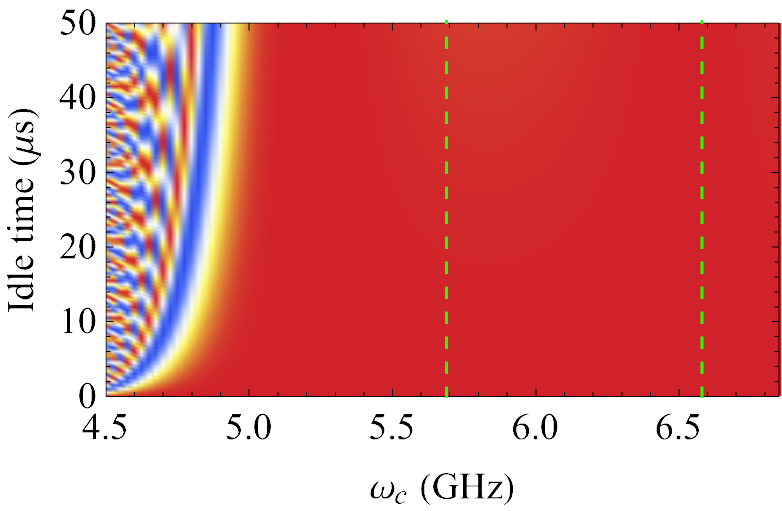}
	\includegraphics[width=0.4\textwidth]{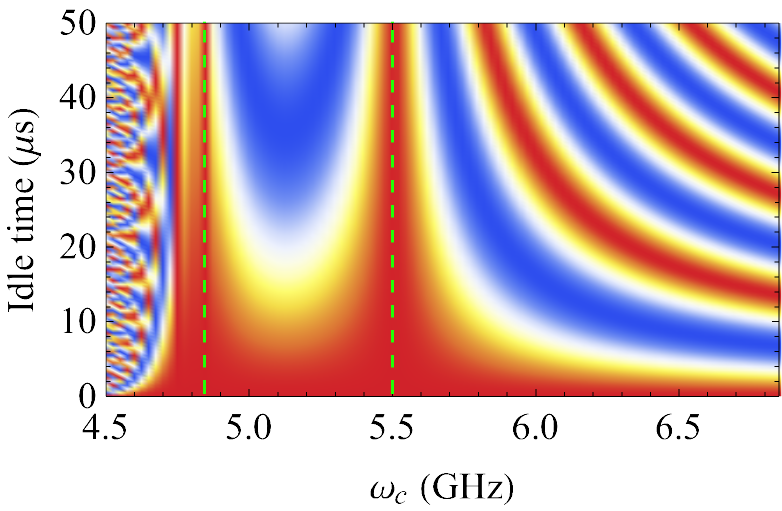}
	\put(-410,130){(c)}\put(-300,40){$\Omega=47$ MHz}
	\put(-205,130){(d)}\put(-140,40){$\Omega=42.3$ MHz}\vspace{-0.3in}\\
	\includegraphics[width=0.4\textwidth]{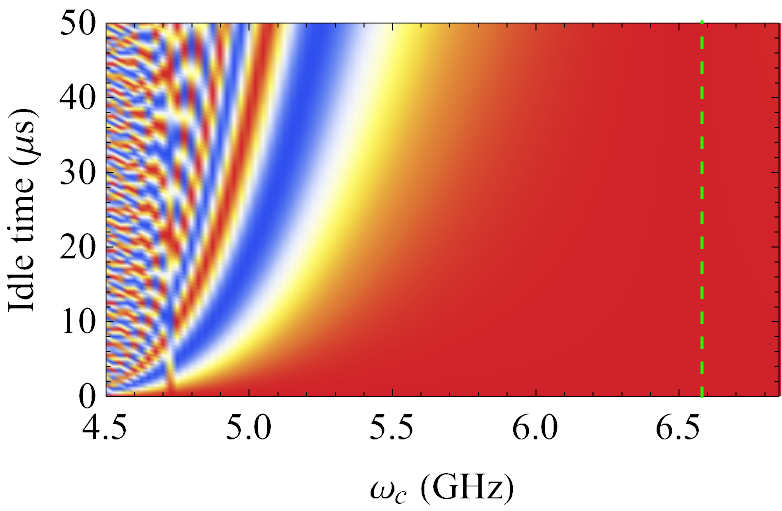}
	\includegraphics[width=0.4\textwidth]{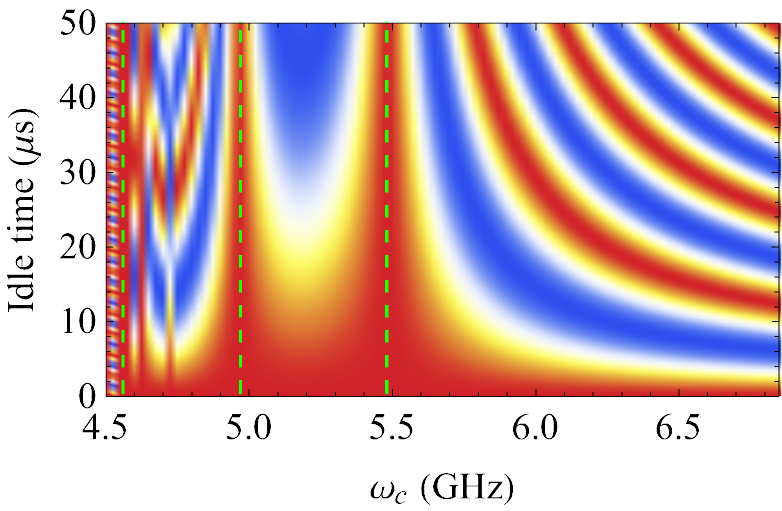}
	\put(-410,130){(e)}\put(-300,40){$\Omega=60$ MHz}
	\put(-205,130){(f)}\put(-140,40){$\Omega=60$ MHz}\\
	%\vspace{-0.1in}
	\caption{Accumulated conditional phase error on computational states on (a,c,e) device 2 at driving amplitude $\Omega=0$, $\Omega=47$~MHz and $\Omega=60$~MHz, and on (b,d,f) device 6 at driving amplitude $\Omega=0$, $\Omega=42.3$~MHz and $\Omega=60$~MHz, respectively. Green dashed lines indicate the $ZZ$-free coupler frequency.}
	\label{fig:pop}
	\vspace{-0.2in}
\end{figure*}
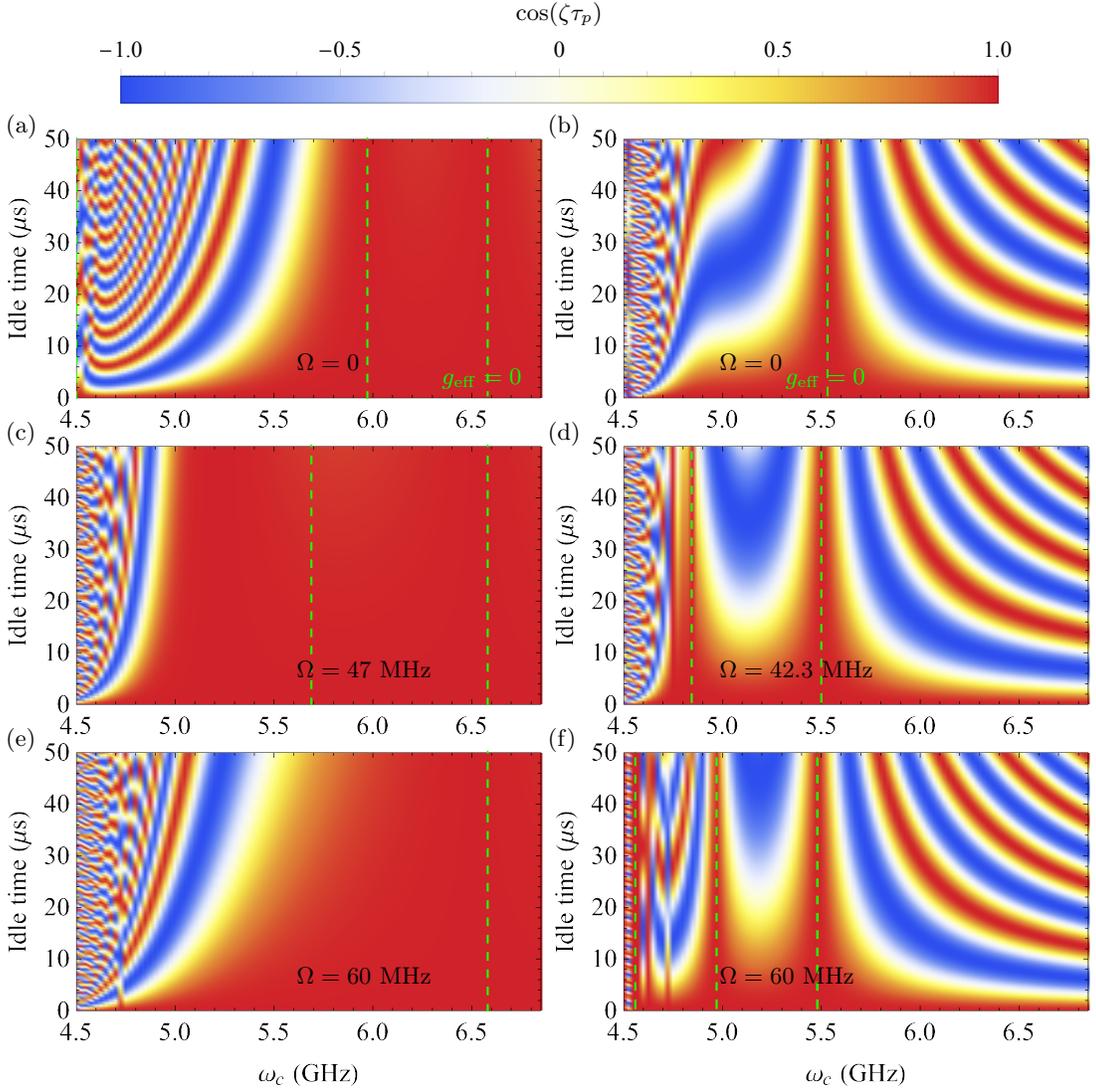

To show how the computational states accumulate conditional phase error, we plot $\exp(i\zeta\tau_p)$ in devices 2 and 6 during the idling periods of duration $\tau_p$ in Fig.~\ref{fig:pop}. Usually such fringes can be measured by performing a Ramsey-like experiment on the target qubit to validate the $ZZ$ cancellation~\cite{ni2021scalable}. 

One can see that device 2 does not accumulate conditional phase at two additional coupler frequencies beyond $g_{\rm eff}=0$ as shown in Fig.~\ref{fig:pop}(a). Figure~\ref{fig:pop}(c) shows that these two $ZZ$-free coupler frequency points reduce to one at the critical amplitude $\Omega=47$ MHz. Above this amplitude, i.e. $\Omega=60$ MHz in Fig.~\ref{fig:pop}(e), the circuit can be free from parasitic $ZZ$ interaction only at $\wia$, indicating that the idle mode is robust against external driving. However, device 6 shows the opposite phenomena. At the idle mode device 6 is only $ZZ$-free at  $\wia$ as shown in Fig.~\ref{fig:pop}(b). When driving amplitude is increased, $ZZ$-freedom can be found in additional coupler frequencies as shown in Fig.~\ref{fig:pop}(d) and \ref{fig:pop}(f).

\vspace{-0.2in}
\section{Impact of Higher order correction}\label{app.higherorder}
\begin{figure}[t]
	\centering
	\includegraphics[width=0.24\textwidth]{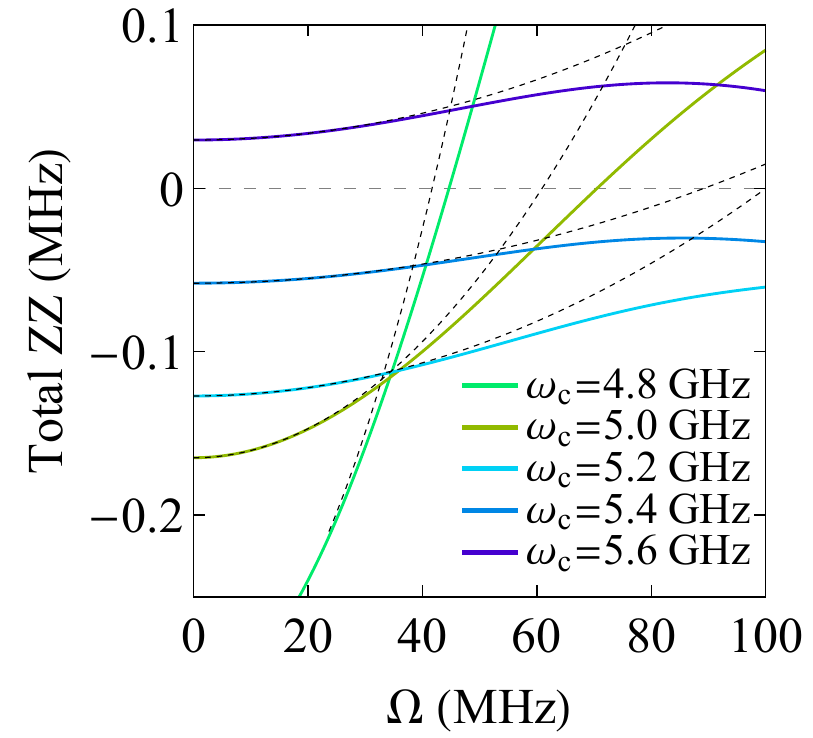}\hspace{-0.1in}
	\includegraphics[width=0.24\textwidth]{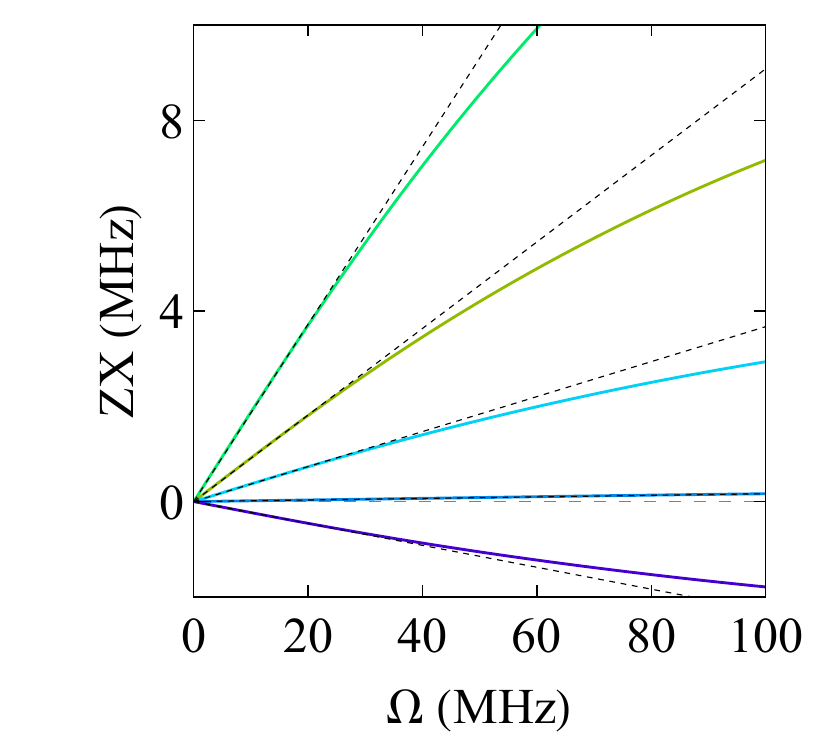}
	\put(-240,105){(a)}\put(-115,105){(b)}\\
	\includegraphics[width=0.24\textwidth]{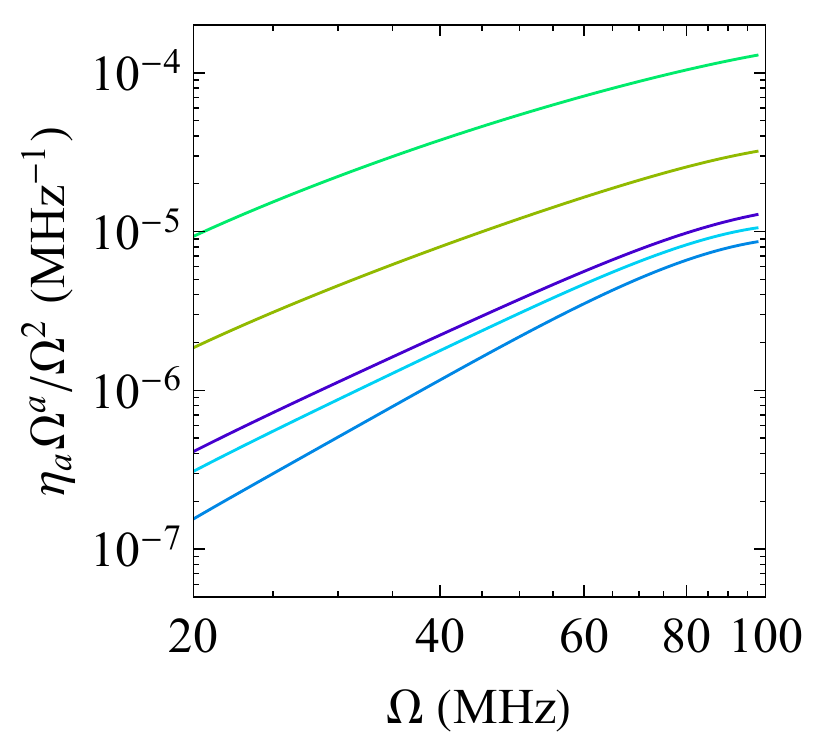}\hspace{-0.1in}
	\includegraphics[width=0.24\textwidth]{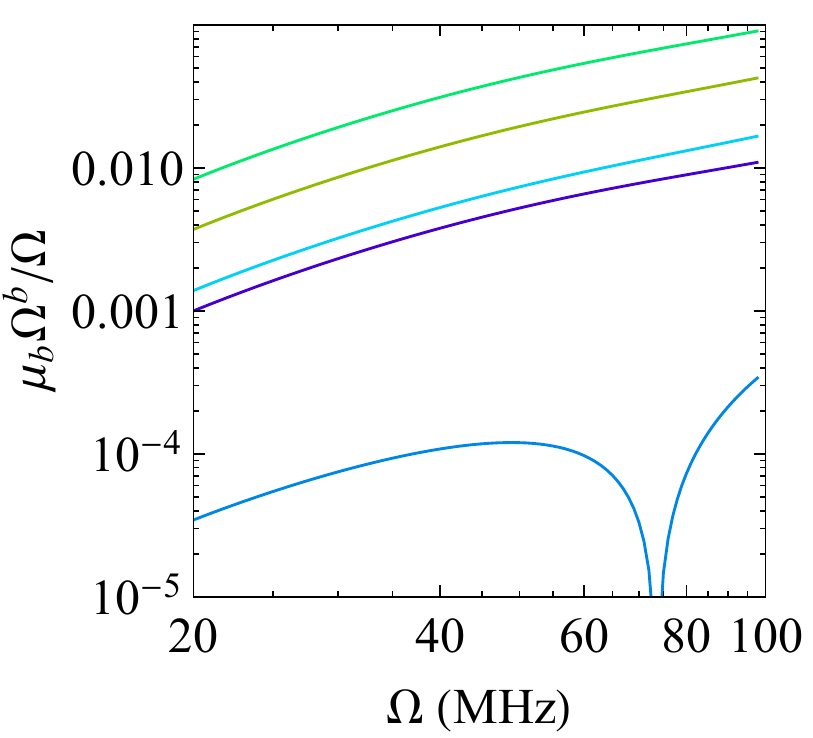}
	\put(-240,105){(c)}\put(-115,105){(d)}\\
	\vspace{-0.1in}
	\caption{In device 6 (a) Total $ZZ$ interaction versus driving amplitude. Dashed lines indicate the trend with respect to only the sum of static and quadratic components (b) $ZX$ rate versus driving amplitude at different coupler frequencies with dashed lines being the linear trend. (c) Higher order correction of $ZZ$ interaction versus driving amplitude on the log-scale.  (d) Higher order correction of $ZX$ rate on the log-scale.}
	\label{fig:hfactor}
\end{figure}

Figure~\ref{fig:hfactor}(a) and~\ref{fig:hfactor}(b) show total $ZZ$ interaction and $ZX$ rate in device 6 at different coupler frequencies. Dashed lines indicate the trend of the Pauli coefficients without higher order correction ($a,b=0$). However in reality corrections from higher levels contribute such that $ZZ$ curves become more flat and finally purely negative with increasing coupling frequency. While $ZX$ rate decreases from positive to negative continuously due to the sign change of $g_{\rm eff}$. The normalized higher order terms are evaluated and plotted in Fig.~\ref{fig:hfactor}(c) and~\ref{fig:hfactor}(d). In the logarithmic scale, these higher-order terms are almost linear at lower driving amplitude, and become more flat with increasing driving amplitude $\Omega$, the slopes also increase with the coupler frequency. Moreover, when the coupler frequency is tuned to be around $\wia$, effective coupling $g_{\rm eff}$ is quite weak and will change its sign. Since $\eta_2\appropto g_{\rm eff}^2$ and $\mu_1\appropto g_{\rm eff}$ are extremely small in the vicinity of the idle coupler frequency $\omega_c^{\rm I}$, higher-order terms contribute dominantly at weak driving amplitude. 

\vspace{0.1in}
\section{Quadratic factor}
\begin{figure}[t!]
	\centering
	\includegraphics[width=0.48\textwidth]{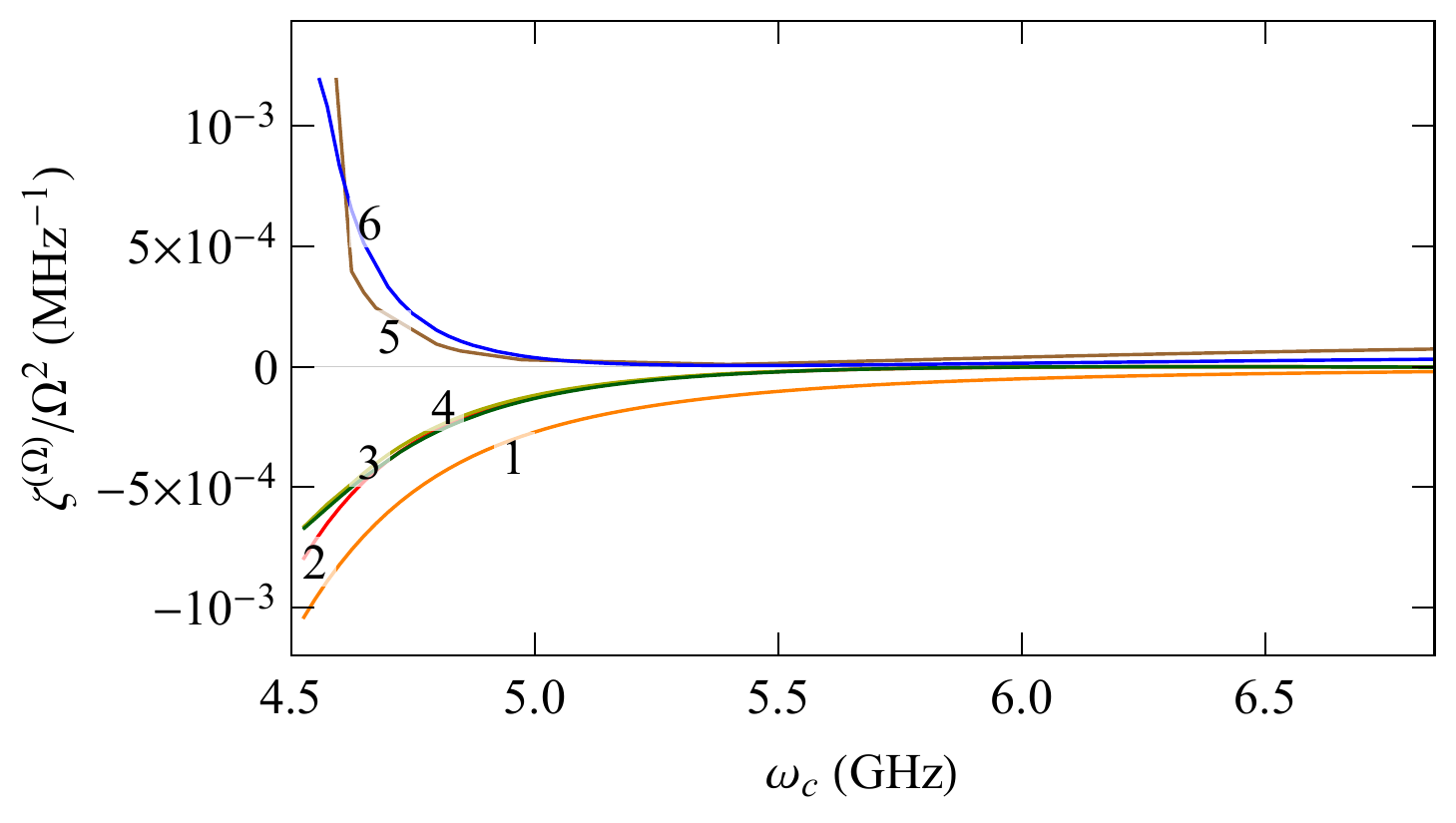}\put(-250,135){(a)}\\
	\vspace{-0.051in}
	\includegraphics[width=0.48\textwidth]{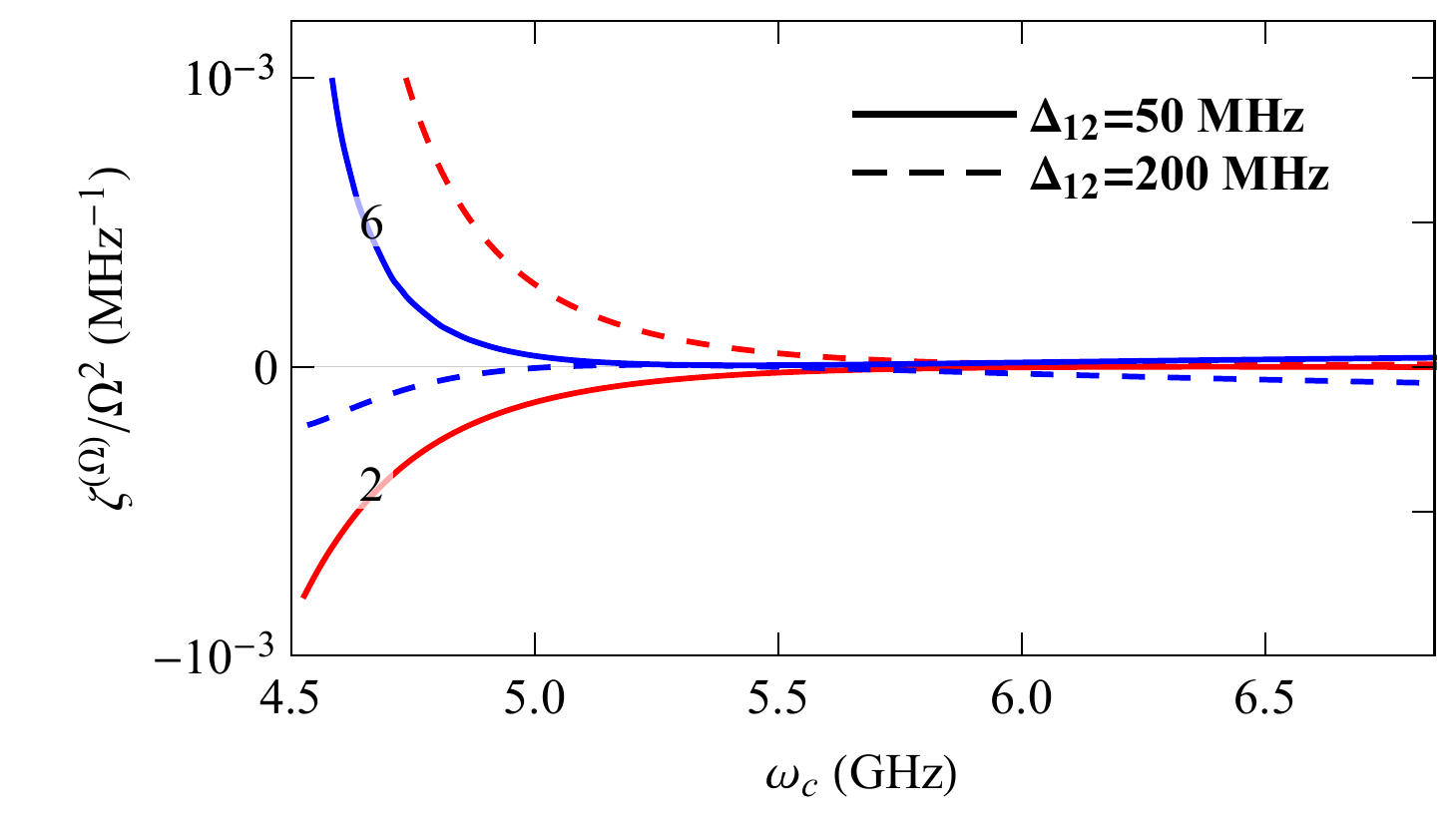}\put(-250,135){(b)}
	\vspace{-0.15in}
	\caption{(a) Quadratic factor $\eta$ as a function of coupler frequency in devices 1-6. (b) Quadratic factor $\eta$ as a function of coupler frequency in devices 2 and 6 at the qubit detuning $\Delta_{12}=50$ MHz and $\Delta_{12}=200$ MHz. }\label{fig:eta}
	\end{figure}

Naively, we can assume the dynamic $ZZ$ interaction is proportional to $\Omega^2$ as the coupler frequency is away from $\wia$. Figure~\ref{fig:eta}(a) shows that such normalized driven part $\zeta_d/\Omega^2$ in device 1-6 dramatically increases when the coupler frequency is close to the qubits, but approaches to zero at higher $\omega_c$. In devices 1-4 the sign of quadratic factor is always negative while in devices 5 and 6 it is positive. Figure~\ref{fig:eta}(b) shows that the sign of dynamic $ZZ$ interaction changes with the qubit-qubit detuning with respect to perturbatively $\zeta_d\appropto1/(\Delta_{12}+\delta_1/2)$.

\bibliography{pfref.bib}
\end{document}